\documentclass[%
rsi,
superscriptaddress,
 amsmath,amssymb,
 reprint,%
 prb,
]{revtex4-1}

\usepackage{graphicx}% Include figure files
\usepackage{dcolumn}% Align table columns on decimal point
\usepackage{bm}% bold math
\usepackage{booktabs}
\usepackage[utf8]{inputenc}
\usepackage[T1]{fontenc}
\usepackage{mathptmx}
\usepackage{etoolbox}
\usepackage{tabularx}
\usepackage{booktabs}
\usepackage{makecell} % optional, for manual line breaks if you want
\usepackage{braket}

\makeatletter
\def\@email#1#2{%
 \endgroup
 \patchcmd{\titleblock@produce}
  {\frontmatter@RRAPformat}
  {\frontmatter@RRAPformat{\produce@RRAP{*#1\href{mailto:#2}{#2}}}\frontmatter@RRAPformat}
  {}{}
}%
\ifdefined\DeclareUnicodeCharacter\DeclareUnicodeCharacter{2212}{-}\fi % guard: undefined under XeTeX/tectonic
\usepackage{hyperref}
\hypersetup{colorlinks,allcolors=blue}

\begin{document}

\title{The qRIXS instrument at the Linac Coherent Light Source}

\author{Lingjia Shen}
\email{lingjias@slac.stanford.edu}
\affiliation{Linac Coherent Light Source, SLAC National Accelerator Laboratory, Menlo Park, CA 94025, USA}%

\author{Daniel Jost}%
\affiliation{Linac Coherent Light Source, SLAC National Accelerator Laboratory, Menlo Park, CA 94025, USA}%

\author{Choongwon Seo}
\affiliation{Linac Coherent Light Source, SLAC National Accelerator Laboratory, Menlo Park, CA 94025, USA}%

\author{Martin Cross}
\affiliation{Linac Coherent Light Source, SLAC National Accelerator Laboratory, Menlo Park, CA 94025, USA}%

\author{Mina R. Bionta}
\affiliation{Linac Coherent Light Source, SLAC National Accelerator Laboratory, Menlo Park, CA 94025, USA}%

\author{Alexander N. Petsch}
\affiliation{Stanford Institute for Materials and Energy Sciences, SLAC National Accelerator Laboratory, Menlo Park, California 94025, USA}

\author{Matthias C. Hoffmann}
\affiliation{Linac Coherent Light Source, SLAC National Accelerator Laboratory, Menlo Park, CA 94025, USA}%

\author{Siqi Cheng}
\affiliation{Linac Coherent Light Source, SLAC National Accelerator Laboratory, Menlo Park, CA 94025, USA}%

\author{Matthew Bain}
\affiliation{Linac Coherent Light Source, SLAC National Accelerator Laboratory, Menlo Park, CA 94025, USA}%

\author{Patrick Oppermann}
\affiliation{Linac Coherent Light Source, SLAC National Accelerator Laboratory, Menlo Park, CA 94025, USA}%

\author{Christina Pino}
\affiliation{Linac Coherent Light Source, SLAC National Accelerator Laboratory, Menlo Park, CA 94025, USA}%

\author{Adam Berges}
\affiliation{Linac Coherent Light Source, SLAC National Accelerator Laboratory, Menlo Park, CA 94025, USA}%

\author{Nicholas Lentz}
\affiliation{Linac Coherent Light Source, SLAC National Accelerator Laboratory, Menlo Park, CA 94025, USA}%

\author{Margaret Ghaly}
\affiliation{Linac Coherent Light Source, SLAC National Accelerator Laboratory, Menlo Park, CA 94025, USA}%

\author{Quentin Yougoubare}
\affiliation{Linac Coherent Light Source, SLAC National Accelerator Laboratory, Menlo Park, CA 94025, USA}%

\author{Frank O'Dowd}
\affiliation{Linac Coherent Light Source, SLAC National Accelerator Laboratory, Menlo Park, CA 94025, USA}%

\author{Jim Defever}
\affiliation{Linac Coherent Light Source, SLAC National Accelerator Laboratory, Menlo Park, CA 94025, USA}%

\author{Amy Liang}
\affiliation{Linac Coherent Light Source, SLAC National Accelerator Laboratory, Menlo Park, CA 94025, USA}%

\author{Eli S. Griff-McMahon}
\affiliation{Linac Coherent Light Source, SLAC National Accelerator Laboratory, Menlo Park, CA 94025, USA}%

\author{Nishant Pandya}
\affiliation{Linac Coherent Light Source, SLAC National Accelerator Laboratory, Menlo Park, CA 94025, USA}%

\author{Sam W. Eisenberg}
\affiliation{Linac Coherent Light Source, SLAC National Accelerator Laboratory, Menlo Park, CA 94025, USA}%

\author{Stephanie Fung}
\affiliation{Linac Coherent Light Source, SLAC National Accelerator Laboratory, Menlo Park, CA 94025, USA}%

\author{Ryan Lodico}
\affiliation{Linac Coherent Light Source, SLAC National Accelerator Laboratory, Menlo Park, CA 94025, USA}%

\author{Joseph Dvorak}
\affiliation{National Synchrotron Light Source II, Brookhaven National Laboratory, Upton, New York 11973, USA}%

\author{Daniele Cocco}
\affiliation{Linac Coherent Light Source, SLAC National Accelerator Laboratory, Menlo Park, CA 94025, USA}%

\author{Joshua J. Turner}
\affiliation{Linac Coherent Light Source, SLAC National Accelerator Laboratory, Menlo Park, CA 94025, USA}%
\affiliation{Stanford Institute for Materials and Energy Sciences, SLAC National Accelerator Laboratory, Menlo Park, California 94025, USA}%

\author{Jake D. Koralek}
\affiliation{Linac Coherent Light Source, SLAC National Accelerator Laboratory, Menlo Park, CA 94025, USA}%

\author{Raybel Almeida}
\affiliation{Linac Coherent Light Source, SLAC National Accelerator Laboratory, Menlo Park, CA 94025, USA}%

\author{K Ninh}
\affiliation{Linac Coherent Light Source, SLAC National Accelerator Laboratory, Menlo Park, CA 94025, USA}%

\author{Douglas Garratt}
\affiliation{Linac Coherent Light Source, SLAC National Accelerator Laboratory, Menlo Park, CA 94025, USA}%

\author{David J. Hoffman}
\affiliation{Linac Coherent Light Source, SLAC National Accelerator Laboratory, Menlo Park, CA 94025, USA}%

\author{Kristjan Kunnus}
\affiliation{Linac Coherent Light Source, SLAC National Accelerator Laboratory, Menlo Park, CA 94025, USA}%

\author{Roopali Kukreja}
\affiliation{Department of Materials Science and Engineering, University of California, Davis, CA, 95616 USA.}%

\author{Alessandra Milloch}
\affiliation{Elettra Sincrotrone Trieste S.C.p.A., Trieste, 34149, Italy.}%
\affiliation{Department of Mathematics and Physics, Università Cattolica del Sacro Cuore, Brescia I-25133, Italy.}%

\author{Christopher T. Parzyck}
\affiliation{Stanford Institute for Materials and Energy Sciences, SLAC National Accelerator Laboratory, Menlo Park, California 94025, USA}%

\author{Wei-Sheng Lee}
\affiliation{Stanford Institute for Materials and Energy Sciences, SLAC National Accelerator Laboratory, Menlo Park, California 94025, USA}%

\author{Antoine Roll}
\affiliation{Condensed Matter Physics and Materials Science Department,
Brookhaven National Laboratory, Upton, New York 11973, USA}%

\author{Mark P. M. Dean}
\affiliation{Condensed Matter Physics and Materials Science Department,
Brookhaven National Laboratory, Upton, New York 11973, USA}%

\author{Sophia F.~R. TenHuisen}
\affiliation{Department of Physics, Harvard University, Cambridge, Massachusetts 02138, USA}%
\affiliation{John A. Paulson School of Engineering and Applied Sciences, Harvard University, Cambridge, Massachusetts 02138, USA}

\author{Matteo Mitrano}
\affiliation{Department of Physics, Harvard University, Cambridge, Massachusetts 02138, USA}%

\author{Thorsten Schmitt}
\affiliation{PSI Center for Photon Sciences, CH-5232 Villigen-PSI, Switzerland}%

\author{Ritika Dagar}
 \affiliation{Stanford PULSE Institute, SLAC National Accelerator Laboratory, Menlo Park, CA 94025, USA}
 \affiliation{Department of Physics, Indian Institute of Information Technology Bhopal, Bhopal, Madhya Pradesh 462003, India}

\author{Shubhadeep Biswas}
\affiliation{Stanford PULSE Institute, SLAC National Accelerator Laboratory, Menlo Park, CA 94025, USA}%
\affiliation{Department of Physics, Indian Institute of Science, Bengaluru-560012, India}

\author{Cheng Peng}
\affiliation{Stanford Institute for Materials and Energy Sciences, SLAC National Accelerator Laboratory, Menlo Park, California 94025, USA}

\author{Joseph Robinson}
\affiliation{Linac Coherent Light Source, SLAC National Accelerator Laboratory, Menlo Park, CA 94025, USA}

\author{Matthias F. Kling}
\affiliation{Linac Coherent Light Source, SLAC National Accelerator Laboratory, Menlo Park, CA 94025, USA}
\affiliation{Stanford PULSE Institute, SLAC National Accelerator Laboratory, Menlo Park, CA 94025, USA}
\affiliation{Department of Applied Physics, Stanford University, Stanford, CA 94305,USA}

\author{Apurva Mehta}
 \affiliation{Linac Coherent Light Source, SLAC National Accelerator Laboratory, Menlo Park, CA 94025, USA}

\author{Giacomo Coslovich}
\email{gcoslovich@slac.stanford.edu}
\affiliation{Linac Coherent Light Source, SLAC National Accelerator Laboratory, Menlo Park, CA 94025, USA}

\author{Georgi L. Dakovski}
\email{dakovski@slac.stanford.edu}
\affiliation{Linac Coherent Light Source, SLAC National Accelerator Laboratory, Menlo Park, CA 94025, USA}

\begin{abstract}
qRIXS is a time-resolved resonant X-ray scattering (RIXS) instrument at the Linac Coherent Light Source (LCLS), primarily designed to exploit the high repetition rate of the LCLS-II X-ray free-electron laser and thereby overcome the low count rates that limit resonant inelastic X-ray scattering (RIXS). It enables high-energy-resolution studies of low-energy excitations in quantum materials driven out of equilibrium, with element and orbital selectivity. Complementary X-ray absorption spectroscopy (XAS) and resonant elastic X-ray scattering (REXS) capabilities support investigations of electronic structure, spin, orbital, and charge order, and their nonequilibrium dynamics. We describe the beam delivery optics and diagnostics, the qRIXS endstation and spectrometer, and the optical pump lasers and timing infrastructure. We report the commissioned performance of the instrument, including a resolving power above 10{,}000 across the soft X-ray window from the O~K to the Cu~L edges at a 33~kHz repetition rate, together with benchmark RIXS spectra of NiO and La$_2$CuO$_4$ and first time-resolved RIXS data on the cuprate superconductor La$_{2-x}$Ba$_x$CuO$_4$. We conclude with an outlook on planned upgrades to improve energy resolution and measurement efficiency and to extend the accessible science.
\end{abstract}

\maketitle

\section{\label{Intro}Introduction}

The Linac Coherent Light Source (LCLS), the world's first hard-X-ray free-electron laser (XFEL), established new opportunities for ultrafast materials-science research. LCLS has also proven to be a versatile source in the soft X-ray regime \cite{Bostedt_rmp_2016}. In particular, the legacy Soft X-ray Research (SXR) instrument provided element- and orbital-selective access to electronic, magnetic, and lattice dynamics on femtosecond timescales. Supported techniques included X-ray absorption spectroscopy (XAS), resonant elastic X-ray scattering (REXS), X-ray photon correlation spectroscopy (XPCS), X-ray magnetic circular dichroism (XMCD), and Fourier-transform holography (FTH) \cite{Heimann-rsi-2011,Schlotter-rsi-2012,Dakovski-jsr-2015,Higley-rsi-2016,Assefa-rsi-2022}. Collectively, these capabilities established resonant soft X-ray experiments at XFELs as a powerful approach for probing quantum matter driven out of equilibrium and for resolving coupled dynamics of spin, charge, orbital, and lattice degrees of freedom with chemical sensitivity and ultrafast time resolution \cite{Graves-NM-2013,Kubacka-science-2014,wang-prl-2012,reid-nc-2018,forst-nm-2015,Johnson-prl-2012,Lee-nc-2012,forst-prl-2014,Iacocca-nc-2019,wandel-science-2022,Tobey-prb-2012,shen-prb-2020,forst-prb-2011,dejong-nm-2013}.

\begin{table}[b]
\caption{\label{tab:1}qRIXS X-ray (high-energy-resolution mode) and laser operation parameters. The spectral brightness refers to the MEG monochromator grating (Table~\ref{tab:2}) performance \cite{nicolas2022}.}
\begin{ruledtabular}
\begin{tabular}{p{0.44\columnwidth}p{0.23\columnwidth}p{0.23\columnwidth}}
Parameter & Designed & \multicolumn{1}{l}{Achieved\footnote{As of December 2025, commissioned at 33~kHz.}} \\
\hline
X-ray energy (eV)                              & 250--1600 & 280--950 \\
X-ray polarization                             & $\pi$ & $\pi$ \\
X-ray spot size ($\mu$m)                       & 10--200 & 10--200 \\
Spectral brightness (ph\,s$^{-1}$\,eV$^{-1}$)  & $10^{15}$ & $10^{15}$ \\
RIXS resolving power ($E/\Delta E$)            & $>10^{4}$ & $>10^{4}$ \\[3pt]
UV/NIR pump (nm)                               & 1030, 515, 343, 257 & 1030, 515, 343, 257 \\
Mid-IR pump ($\mu$m), tunable                  & 1.3--17 & 1.3--15 \\
THz pump (THz)                                 & 1--3 & --- \\
\end{tabular}
\end{ruledtabular}
\end{table}

Resonant inelastic X-ray scattering (RIXS) is a photon-in/photon-out spectroscopy in which an incident X-ray photon of energy \(E_i=\hbar\omega_i\) and wave vector \(\mathbf{k}_i\) is tuned near an element-specific absorption edge. Absorption promotes a core electron into an unoccupied state and creates a short-lived intermediate state containing a core hole. Radiative decay of this intermediate state produces an emitted photon of energy \(E_f=\hbar\omega_f\) and wave vector \(\mathbf{k}_f\). If the material does not return to its initial state, the emitted photon has exchanged energy and momentum with the sample. The measured transfers are
\begin{equation}
  \hbar\omega = E_i-E_f,
  \qquad
  \hbar\mathbf{q}=\hbar(\mathbf{k}_i-\mathbf{k}_f),
\end{equation}
where \(\hbar\omega\) is the energy transferred to the material, positive for energy loss, and \(\hbar\mathbf{q}\) is the momentum transfer. By measuring the emitted photons as a function of \(\hbar\omega\) and \(\hbar\mathbf{q}\), RIXS therefore maps the dispersion of the elementary excitations of the material.

A distinctive strength of RIXS as a technique is its sensitivity to multiple types of elementary excitations. Depending on the absorption edge, scattering geometry, polarization, and energy range, RIXS can probe lattice vibrations, spin excitations, orbital and crystal-field excitations, and a range of charge excitations including plasmons, charge-transfer excitations, excitons, and electron--hole continua~\cite{Ament-rmp-2011}. This broad sensitivity arises from the resonant second-order scattering process, in which the core-excited intermediate state couples to valence degrees of freedom through core-hole, spin-orbit, and multiplet interactions. Selection of the absorption edge provides element and, in favorable cases, site specificity, while the scattering geometry and incident and emitted polarizations modify the relative spectral weights of charge, spin, orbital, and lattice channels. RIXS can therefore provide a unified, momentum-resolved view of coupled electronic and structural excitations over energy scales ranging from millielectronvolts to electronvolts~\cite{Ament-rmp-2011,Mitrano-prx-2024}.

\begin{figure*}[t]
  \centering
  \includegraphics[width=0.9\linewidth]{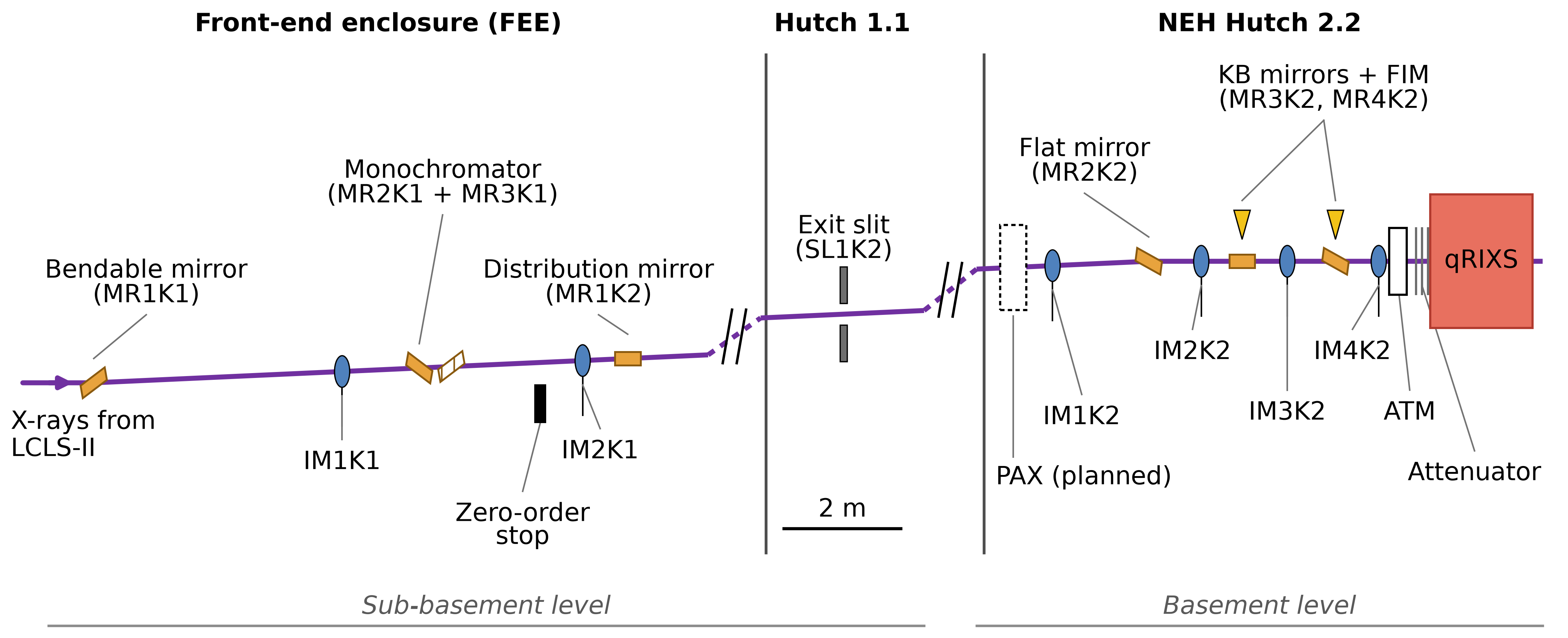}
  \caption{Functional layout of the RIX beamline from the front-end enclosure (FEE) to qRIXS in NEH Hutch 2.2 (side view). Components are to scale within each area; drifts between areas are compressed at the break marks. FEE and Hutch 1.1 are at the sub-basement level, Hutch 2.2 at the basement level. MR1K1 deflects the beam upward into the monochromator (plane pre-mirror MR2K1 and VLS grating MR3K1), MR1K2 deflects it horizontally into the RIX branch, and MR2K2 returns it to horizontal before the KB pair MR3K2 (horizontal) and MR4K2 (vertical) focuses it onto the sample. IM: YAG profile monitors on insertion paddles; FIM: fluorescence intensity monitors in the KB tanks; ATM: X-ray arrival time monitor; PAX (dashed): planned photoelectron spectrometer for incident X-ray energy monitoring \cite{Higley-jsr-2022}.}
  \label{fig:1}
\end{figure*}

Time-resolved RIXS (tr-RIXS) extends these strengths into the nonequilibrium regime by tracking elementary excitations following an ultrafast perturbation. It provides a momentum-resolved view of how different degrees of freedom evolve on timescales set by their underlying interactions. In quantum materials, where competing interactions and intertwined orders govern phenomena such as superconductivity, charge- and spin-density waves, orbital order, and topological magnetism, tr-RIXS can help establish microscopic cause-and-effect relationships by monitoring spectral changes as a system is driven across a photoinduced phase transition or into a metastable state~\cite{Mitrano-prx-2024}. In functional materials and devices, including batteries, microelectronics, and oxide or van der Waals heterostructures, performance is often governed by nanoscale interfaces, strain fields, and nonequilibrium carrier distributions~\cite{piro-nrm-2021,sakizadeh-arpc-2025}. In these systems, tr-RIXS provides element- and site-selective sensitivity to interfacial electronic structure and excitations, enabling direct tests of how specific materials-design parameters affect collective modes and dissipation channels~\cite{degroot-nrmp-2024}. Motivated by this potential, pioneering tr-RIXS measurements were carried out at SXR~\cite{mitrano-sa-2019,Mitrano-prb-2019,paris-npjqm-2021,Parchenko-prr-2020,lu-sr-2020}, and dedicated tr-RIXS instruments have since been constructed at XFEL facilities worldwide~\cite{furka,hrixs}.

These early experiments demonstrated the scientific potential of tr-RIXS, but they were limited by the repetition rate of the copper linac, which delivered X-ray pulses at up to 120~Hz~\cite{Emma-np-2010}. The resulting limit on average photon flux is particularly severe for RIXS because, despite the resonant enhancement, the detected signal is intrinsically weak. At the $L$ edges of the $3d$ transition metals only a small fraction of the core holes decays by emitting a photon; the rest decay through Auger and Coster--Kronig processes~\cite{Krause-jpcrd-1979}. Of the photons that are emitted, only those that fall within the small solid angle of the spectrometer, are diffracted into the selected order and are registered by the detector contribute to the spectrum. Consequently, only a minute fraction of the incident photons is detected as inelastic signal, and a high-resolution RIXS spectrum at 120~Hz requires hours of acquisition. The superconducting linac of LCLS-II~\cite{dunne-spie-2023} raises the available repetition rate by several orders of magnitude, ultimately to 1~MHz, and thereby directly addresses this central limitation, but exploiting it requires a dedicated high-resolution instrument with pump--probe capabilities. We therefore developed qRIXS. In this article, we describe the instrument and its supporting infrastructure and summarize the initial performance demonstrated during its first year of operation (Table~\ref{tab:1}).

\section{\label{RIX}RIX Beamline}
\subsection{X-ray transport}

In this section, we briefly describe the layout of the RIX beamline, which extends from the front-end enclosure (FEE) through Hutch~1.1 to Hutch~2.2 of the Near-Experiment Hall (NEH), where the qRIXS instrument is located (Fig.~\ref{fig:1}). The LCLS-II soft X-ray beam enters the FEE at the NEH sub-basement level, where a bendable elliptical mirror (MR1K1) deflects it upward and establishes the convergence required by the monochromator. The beam then passes through a variable-line-spacing (VLS) plane-grating monochromator operating in converging light~\cite{nicolas2022}, comprising a plane pre-mirror (MR2K1), which sets the included angle at the grating, and a plane grating (MR3K1), which disperses the beam in the vertical plane. The arrangement follows the converging-light VLS geometry of Hettrick and Underwood~\cite{Hettrick-aip-1986}, in which the groove-density variation of a plane grating provides the focusing, combined with the plane pre-mirror of the plane-grating monochromator of Petersen~\cite{Petersen-oc-1982}, which allows the included angle to be varied. The VLS groove profile, together with the convergence provided by MR1K1, keeps the dispersed beam focused on the exit slit in Hutch~1.1, which selects the photon-energy bandwidth transmitted downstream; a zero-order stop behind the grating blocks the undispersed beam. Interchangeable gratings provide access to different photon-energy ranges and resolving powers. Each grating is operated at a fixed-focus constant \(C_{\mathrm{ff}} = \cos\beta/\cos\alpha\), where \(\alpha\) and \(\beta\) are the angles of incidence and diffraction at the grating, which sets the trade-off between resolving power, transmission, and pulse stretching. Their specifications and operating \(C_{\mathrm{ff}}\) values are summarized in Table~\ref{tab:2}; the calculated resolving power and transmitted pulse energy of each grating as a function of photon energy, together with the resulting pulse stretching, have been reported in Ref.~\onlinecite{nicolas2022}.

\begin{table}[tb]
\centering
\caption{qRIXS high-energy-resolution monochromator gratings.}
\label{tab:2}
\renewcommand{\arraystretch}{1.15}
\setlength{\tabcolsep}{6pt}

\begin{tabular}{@{}c c c c@{}}
\toprule
\hline
\textbf{Grating} &
\makecell{\textbf{Line density}\textbf{(l/mm)}} &
\textbf{$C_{ff}$} &
\makecell{\textbf{Operation} \textbf{window (eV)}} \\
\hline

\midrule
HRG & 700 & 3.0 & 670--1500 \\
MEG & 450 & 3.2 & 415--1110 \\
LEG & 300 & 3.2 & 274--740 \\
\bottomrule
\hline
\end{tabular}
\end{table}

Downstream of the monochromator, the distribution mirror MR1K2 deflects the beam horizontally into the RIX branch. After the exit slit, the diverging beam rises to the NEH basement level, where Hutch~2.2 is located, and is intercepted by the flat mirror MR2K2, which returns the trajectory to the horizontal plane. Final focusing onto the qRIXS interaction point (IP) is provided by the Kirkpatrick--Baez (KB) mirror pair MR3K2 and MR4K2, which focus in the horizontal and vertical directions, respectively, and also allow small trajectory adjustments in \(x\) and \(y\) before the beam enters the qRIXS chamber. Downstream of qRIXS, the beam can be transmitted to the chemRIXS instrument for solution-phase research~\cite{chemrixs}, which shares the upstream RIX beamline components with qRIXS and is omitted from Fig.~\ref{fig:1} for simplicity.

The Experimental Physics and Industrial Control System (EPICS) provides the underlying control infrastructure for the RIX beamline and the qRIXS endstation~\cite{epics_overview,epics_intro}. EPICS implements a distributed, network-based architecture in which process variables represent device states and are accessed via standardized interfaces, enabling robust communication between hardware components, embedded controllers, and user-level applications. This design supports scalable integration of heterogeneous subsystems and facilitates coordinated operation across the full instrument.

We use EPICS throughout the beamline and the endstation, ensuring a unified control environment. This allows synchronized control of optical elements, diagnostics, sample environments, and detectors, while also providing a common framework for automation, monitoring, and data acquisition. The use of EPICS therefore contributes to reliable operation and reproducible experimental conditions in complex soft X-ray spectroscopy measurements~\cite{epics_recent}.

\subsection{Beamline detectors and diagnostics}

\begin{figure}[t]
  \centering
  \includegraphics[width=0.95\linewidth]{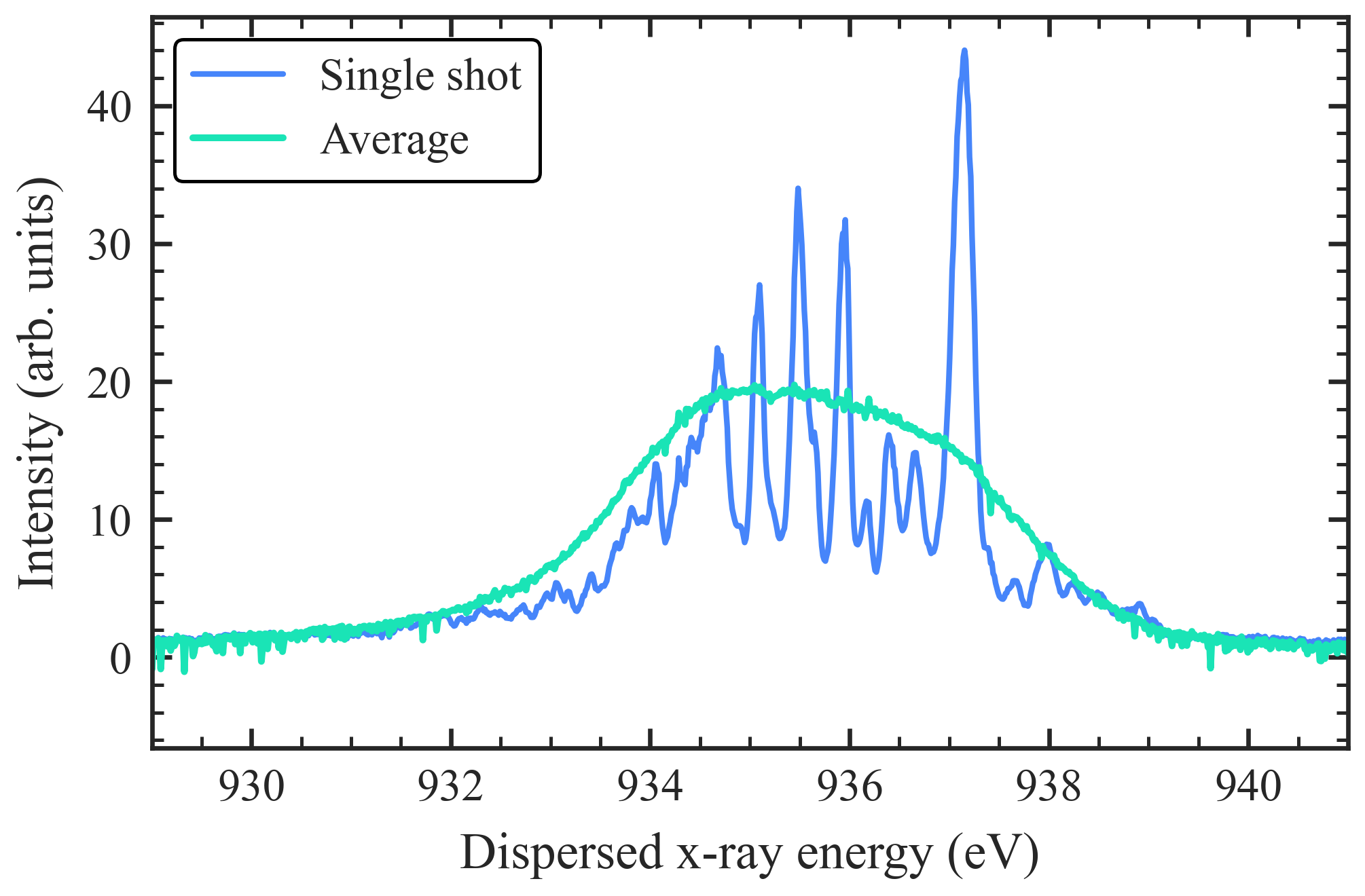}
  \caption{Self-amplified spontaneous emission spectrum of the LCLS-II X-ray beam dispersed by the medium-energy high-energy-resolution grating (MEG) and measured with the SL1K2 spectrometer.} %The spectrum is averaged over 6986 XFEL pulses.}
  \label{fig:2}
\end{figure}

A set of detectors and diagnostics are installed along the beam path for alignment, transmission verification, and shot-to-shot monitoring (Fig.~\ref{fig:1}). Diagnostics include X-ray profile monitor (e.g. IM1K1) based on a yttrium aluminum garnet (YAG) scintillator imaged by an optical camera. These monitors are mounted on insertion paddles; the same paddles can also host an X-ray power meter \cite{Heimann_2025}, enabling straightforward measurements of beam transmission through individual beamline sections. We use these diagnostics to establish and maintain a pre-calibrated beam trajectory and to verify that the measured transmission remains consistent with nominal performance. The diagnostic readbacks are available to the online analysis pipeline, enabling efficient alignment and rapid performance checks.

\begin{figure*}[t]
  \centering
  \includegraphics[width=0.9\linewidth]{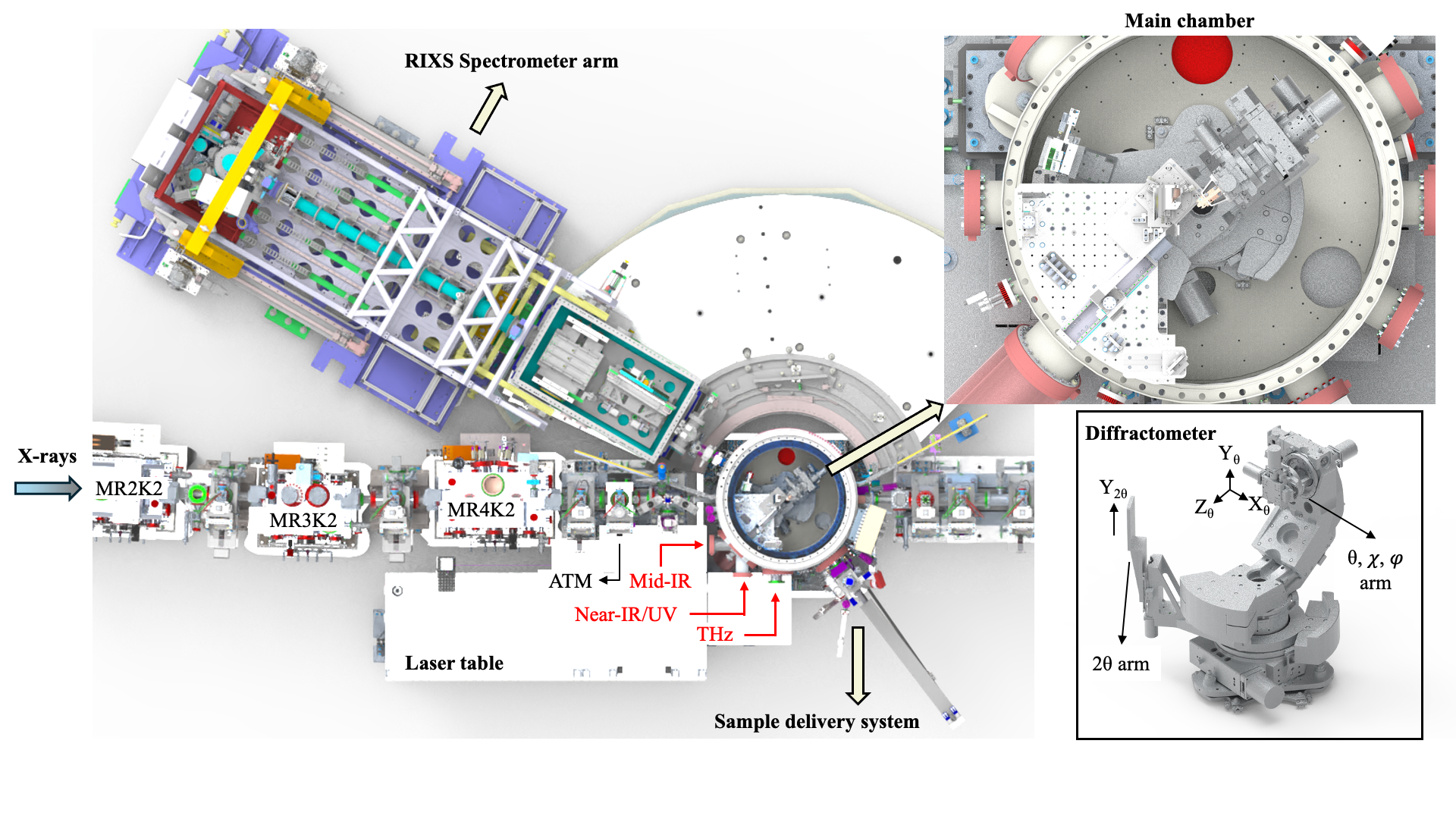}
  \caption{Major components of the qRIXS instrument (top view). The X-ray beam enters from the left and passes the flat mirror MR2K2, the Kirkpatrick--Baez pair MR3K2 and MR4K2 and the X-ray arrival time monitor (ATM) before it reaches the main chamber; the RIXS spectrometer arm, with the collection mirror, gratings and detector, extends to the upper left at the scattering angle $2\theta$. The laser table below the beamline delivers the near-infrared/UV, mid-infrared and THz pump beams to the chamber through the in-coupling ports marked by the red arrows, and the sample delivery system is attached at the lower right. Upper inset: interior of the main chamber with the in-vacuum breadboard and the diffractometer. Lower inset: the diffractometer with the sample translations.}
  \label{fig:3}
\end{figure*}

The first dedicated detector is the spectrometer SL1K2 at the exit-slit location, which consists of a YAG screen viewed by a CMOS camera (Manta G419B)\cite{manta_g419b_datasheet}. It provides a direct measurement of the vertically dispersed self-amplified spontaneous emission (SASE) spectrum. This is shown in \autoref{fig:2}. The single-shot spectrum exhibits a spiky structure due to the stochastic amplification of shot noise in the electron bunch entering the undulator, resulting in multiple independent longitudinal coherence modes within the FEL gain bandwidth \cite{Bostedt_rmp_2016}. Downstream of SL1K2, two fluorescence intensity monitors (FIMs) \cite{heimann2019fim,Heimann_2025} positioned inside the KB optic tanks (MR3K2 and MR4K2) measure the incident X-ray flux for normalization. Between MR4K2 and the qRIXS chamber, an X-ray arrival-time monitor (ATM) tracks shot-to-shot timing jitter of the XFEL pulses; this system is discussed further below.

\section{\label{qRIX}qRIXS Instrument}
As shown in \autoref{fig:3}, the qRIXS instrument consists of the main chamber, the RIXS spectrometer arm, the sample delivery system (SDS), and the pump laser system. In this section, we describe the main chamber, the RIXS spectrometer arm, and the SDS; the pump laser system is discussed in the next section.

\subsection{Main chamber}

The qRIXS main chamber is a single cylindrical ultra-high vacuum (UHV) vessel, approximately 1~m in diameter and routinely operated at pressures in the \(10^{-10}\)~Torr range. Centered on the interaction point, it provides the mechanical support and common vacuum environment for the endstation. The incident X-ray beam enters the chamber horizontally from the west side. During RIXS operation, the sample intercepts the beam and the scattered photons are directed into the RIXS spectrometer through a rotatable exit port (\autoref{fig:3}). Alternatively, with the diffractometer retracted, the beam passes through the chamber and exits on the east side for transmission to the downstream chemRIXS endstation~\cite{chemrixs}. To maintain continuous angular coverage in the horizontal scattering plane, the exit port is realized with a sliding-seal intermediate section (CINEL~\cite{cinel}) that permits motion from approximately \(40^\circ\) to \(150^\circ\) without breaking vacuum.

The chamber is organized into three principal assemblies: 1) a lower chamber mounted on a granite base, 2) an intermediate chamber that incorporates the beam entrance and the rotating scattered-beam exit path, and 3) a removable top plate that carries viewports, service ports, and a large central access opening. The section below the chamber provides the primary mounting interfaces for the diffractometer and other internal support structures, and also accommodates the pumping hardware and electrical feedthroughs required for operation. The top section, by contrast, provides direct line-of-sight access to the interaction region for alignment, visual inspection, sample handling, and cryogenic integration. This three-part architecture gives the chamber a modular and serviceable layout while preserving mechanical rigidity and UHV compatibility. It also leaves dedicated internal volume for auxiliary in-vacuum instrumentation, including the breadboard-supported assembly described in the following subsection.

\subsubsection{Diffractometer and cryostat}
The main chamber houses an in-vacuum diffractometer that allows precise and reproducible alignment of the sample with respect to the incident X-ray beam, optical pump, and detectors. In the qRIXS geometry, the interaction point is defined by the spatial overlap of the sample, X-ray beam, and optical pump, and must be positioned at the center of rotation of the diffractometer. This requirement places constraints on positional stability throughout alignment and measurement. 

The diffractometer (Huber~\cite{huber}) is operated in a horizontal scattering geometry and combines multi-axis sample manipulation with compatibility for cryogenic sample environments and SLAC Beckhoff-based motion control~\cite{beckhoff}. The system provides eight motorized degrees of freedom: three sample translations (X$_{\theta}$, Y$_{\theta}$, Z$_{\theta}$) stacked on the \(\theta\) rotation stage, which bring the sample surface to the center of rotation; a vertical translation of the in-vacuum detector arm (Y$_{2\theta}$); and four rotations (\(\theta\), \(2\theta\), \(\chi\), and \(\phi\)). All axes are located inside the vacuum chamber (Fig.~\ref{fig:3}), so that alignment of the sample to the X-ray beam and to the rotation center is performed without moving the chamber itself. The \(\theta\) and \(2\theta\) axes are concentric and define the vertical rotation axis of the diffractometer. The extended \(2\theta\) arm carries an in-vacuum detector, an avalanche photodiode (Laser Components~\cite{lasercomponents}) or an area detector, whose vertical translation (Y$_{2\theta}$) serves alignment and reciprocal-space surveys; this detector provides the XAS and resonant REXS measurements that complement RIXS, for example for locating resonances and ordering wavevectors before a RIXS scan. The motion performance includes translational resolution of 0.5~\(\mu\)m, angular resolution down to 0.001\(^\circ\), translational repeatability of 2~\(\mu\)m, and a sphere of confusion below 45~\(\mu\)m. The mechanical stability exceeds 65 Hz in first natural frequency, and the dimensional stability is specified to remain within \(\pm 10~\mu\)m for a \(\pm 1\,^\circ\)C temperature variation. Together, these characteristics enable reliable low-temperature, momentum-resolved qRIXS measurements with reproducible sample positioning and scattering geometry.

The diffractometer is integrated with a top-loading helium-flow cryostat that provides cryogenic temperatures down to approximately 20\,K. The sample is thermally coupled to the cryostat through a flexible copper braid, which provides high thermal conductance while accommodating diffractometer motion. The translational and rotational motions of the diffractometer and cryostat are coordinated to ensure safe operation while maintaining stable cryogenic performance.

\subsubsection{In-vacuum breadboard}

\begin{figure}[b]
  \centering
  \includegraphics[width=0.95\linewidth]{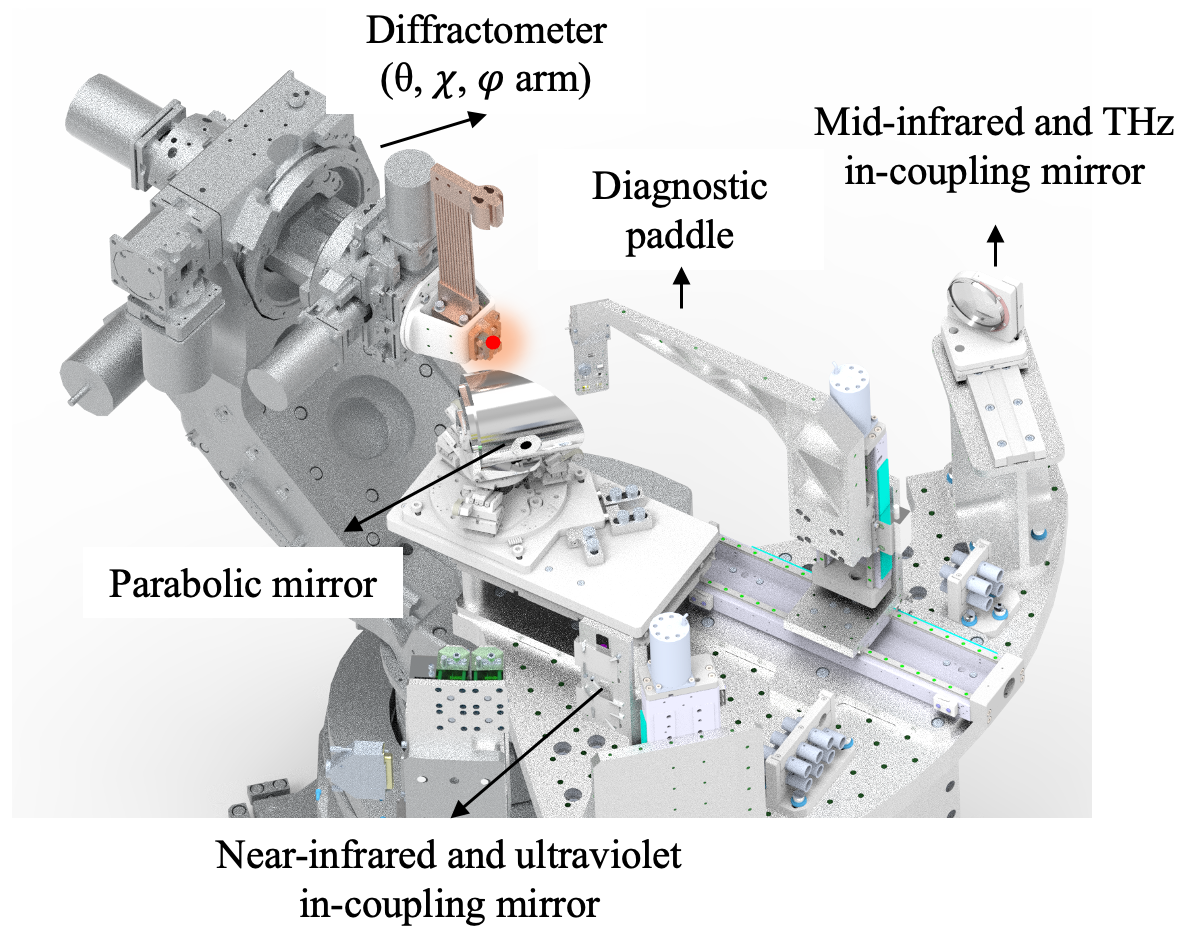}
  \caption{Layout of in-vacuum breadboard with respect to the diffractometer. The red dot is the interaction point.}
  \label{fig:4}
\end{figure}

An in-vacuum breadboard mounted in the southwest quadrant of the qRIXS main chamber serves as the optical bridge between the pump-laser in-coupling ports and the interaction point. This assembly supports the relay and alignment optics required to transport optical pump beams from the chamber-side laser entry ports to the sample position (\autoref{fig:3}). In the present implementation, the optical path is defined by steering mirrors that provide the flexibility needed to align the pump beam within the chamber volume. For mid-infrared and THz excitation, a 90\textdegree{} off-axis parabolic mirror with a 76.2~mm effective focal length focuses the pump beam onto the interaction point. The optics are carried by vacuum-compatible motorized stages. The visible-beam optics and the diagnostic paddle are moved by linear piezo stages with closed-loop linear encoders (PPS-60-SM, Micronix~\cite{micronix}), driven by the Beckhoff motion controllers of the instrument~\cite{beckhoff}; the mid-infrared optics by a linear stage and a tip-tilt stage (SLLV472 and STT-50.8-UHV, SmarAct~\cite{smaract}); and the THz optics by a six-axis hexapod (SP~110.45.1-UHV, SmarAct) and two linear stages (SLC-24120-UHV, SmarAct), all with closed-loop encoders except the open-loop tip-tilt stage. A further SmarAct stage with an absolute encoder (SLC-1730-UHV) provides the horizontal motion of the diagnostics. The hexapod, the THz translation and the diagnostic stages are interlocked with the diffractometer to prevent collisions.

The breadboard also houses a retractable diagnostic paddle used for both X-ray and pump laser alignment and characterization at the interaction point, including timing-related measurements and knife-edge scans (\autoref{fig:4}). By concentrating these functions on a dedicated in-vacuum support structure, the breadboard provides a modular platform for pump delivery and beam diagnostics while preserving clear X-ray access to the diffractometer. This arrangement simplifies alignment and commissioning of optical pump configurations and allows diagnostic tools to be introduced or retracted without disturbing the core scattering geometry.

\subsection{Sample delivery system}

\begin{figure}[b]
  \centering
  \includegraphics[width=0.95\linewidth]{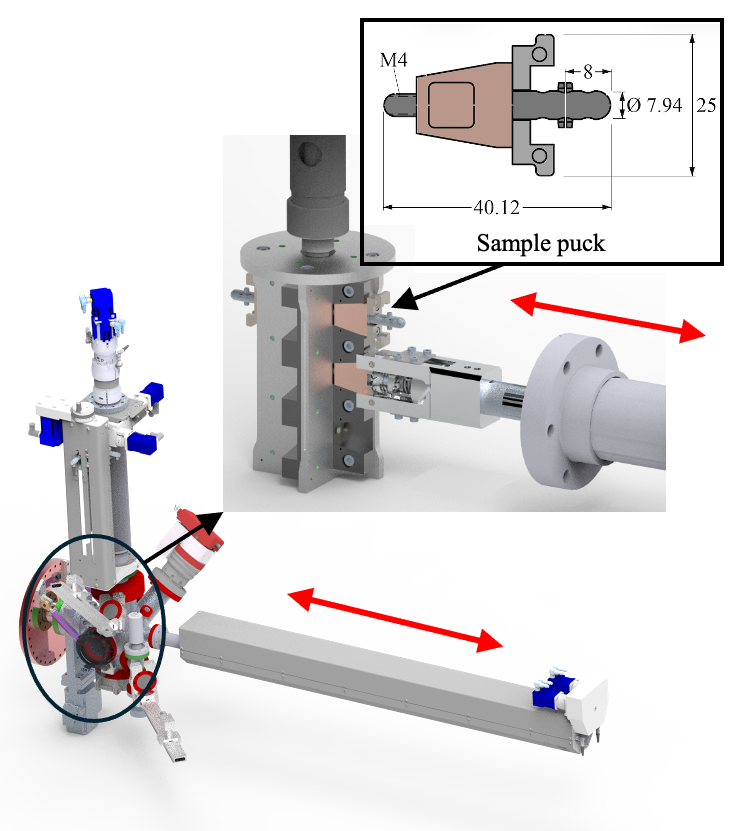}
  \caption{The qRIXS sample delivery system. Red arrows mark the transfer direction into the main chamber. Inset: In-vacuum layout of the sample garage, with a sample puck engaged by the transfer arm. The boxed drawing shows a side view of the sample puck, consisting of a copper shuttle (brown) and a retainer plate with the sample screw (gray); dimensions are in mm.}
  \label{fig:5}
\end{figure}

The qRIXS SDS is designed to enable rapid exchange of solid samples under UHV while preserving the cryogenic sample environment at the interaction point. The system consists of a demountable sample puck, a motorized load-lock transfer assembly, and a cryogenic sample mount thermally coupled to the cryostat. Samples are mounted on wedge-shaped copper pucks with a captive central screw, allowing repeated mounting and removal while the screw clamps the puck firmly against the sample holder for reliable mechanical and thermal contact (\autoref{fig:5}). Samples are introduced through a dedicated load-lock chamber connected to the main chamber through a gate valve on the southeast side (\autoref{fig:3}), and are transferred in vacuum by a linear-rotary transfer arm. A vertically actuated rotary carousel inside the load lock stores up to 12 sample pucks, so that multiple samples can be loaded in air and exchanged without venting the main chamber. The load lock is pumped to a nominal base pressure of \(10^{-8}\)~Torr level before the gate valve to the main chamber is opened. During mounting, the transfer arm engages the threaded puck, inserts it onto the sample holder, and applies sufficient torque to secure reliable positioning and thermal contact. The sample mount is connected to the cryostat cold finger through a flexible oxygen-free high-conductivity copper braid, while a rigid support structure maintains mechanical stability. Together, these features provide a compact UHV-compatible architecture for efficient sample exchange and low-temperature qRIXS measurements. 

\subsection{qRIXS spectrometer}
\subsubsection{Optical scheme}\label{sec:optic_scheme}

The qRIXS spectrometer is designed to provide both a high-resolution mode for precision measurements and a higher-throughput, medium-resolution mode for more time-efficient surveys of parameter space. To minimize alignment complexity at an XFEL endstation, we restricted the optical layout to a simple spherical variable-line-spacing (SVLS) grating geometry with a fixed source-to-grating distance and only a single optical element in the vertical dispersive plane. We then added a horizontally deflecting parabolic collection mirror to improve throughput. In addition to increasing collection efficiency, the horizontally collimated beam suppresses curvature of the footprint at the detector in the non-dispersive direction and preserves a future path toward polarization analysis. High resolution is achieved with a high-line-density grating operated in positive diffraction orders up to $m=4$. By varying the included angle at the grating, the spectrometer can be tuned to zero-coma conditions across the operating range, thereby enabling the full useful acceptance of a 300~mm grating to be utilized.

Among the configurations evaluated during the preliminary optical design, we have chosen a spherical-grating architecture because it meets both constraints with a single optic in the dispersive plane, i.e. optical simplicity for rapid alignment, and a fixed grating position. We selected positive diffraction order over negative diffraction order because it more readily satisfies the resolution requirement for the relatively large vertical source size needed to avoid sample damage by intense XFEL illumination. It also provides substantially better tuning behavior across the operating range. The use of an additional pre- or post-mirror to maintain a fixed output-beam direction was also considered. A premirror was excluded by space constraints associated with the horizontal collection mirror, whereas a postmirror was rejected because the added reflectivity loss and degradation in resolving power outweighed the benefit of a fixed output direction. Finally, downward deflection was chosen so that the detector remains closer to the floor, which is advantageous for mechanical stability.

\begin{table}[t]
\caption{\label{tab:svls_grating_parameters}Parameters of the two blazed SVLS gratings: $r_a$ is the source-to-grating distance, $\alpha$ is the grazing angle of incidence measured from the grating normal at the pseudo-Rowland condition, $R$ is the radius of curvature of the spherical substrate, and $k_0$ is the central groove density. The local groove density varies along the dispersive coordinate $w$ on the grating surface as $k(w) = k_0 + a_1 w + a_2 w^2 + a_3 w^3$, with $w = 0$ at the grating center. }
\begin{ruledtabular}
\begin{tabular}{lcc}
Parameter & High-resolution & Medium-resolution \\
\hline
$r_a$ (mm) & 2200 & 2200 \\
$\alpha$ (pseudo-Rowland) (deg) & 88.5 & 88.0 \\
$k_0$ (mm$^{-1}$) & 3600 & 1500 \\
$a_1$ (mm$^{-2}$) & 1.6003 & 0.5400 \\
$a_2$ (mm$^{-3}$) & $6.818\times10^{-4}$ & $2.431\times10^{-4}$ \\
$a_3$ (mm$^{-4}$) & $2.70\times10^{-7}$ & $1.21\times10^{-7}$ \\
$R$ (mm) & 84043.4 & 63038.2 \\
\end{tabular}
\end{ruledtabular}
\end{table}

The optical parameters were optimized by balancing resolving power, throughput, detector travel, and focal-plane geometry within the available mechanical envelope. The principal free parameters were the grating incidence angle $\alpha$, source-to-grating distance $r_a$, central groove density $k_0$, radius of curvature $R$, and the SVLS coefficients. The optimization was referenced to 930~eV, corresponding to the Cu $L_3$ edge. Because FEL damage considerations require a vertical beam size of approximately $20~\mu$m FWHM at the sample, the design was optimized under a constant power-density constraint. Under this condition, the most efficient high-resolution solution was found to use a 3600~mm$^{-1}$ grating at $r_a = 2200$~mm. For the medium-resolution mode, the grating parameters were chosen not only to improve efficiency, but also to keep the detector trajectory reasonably close to that of the high-resolution mode, thereby simplifying the mechanical design. The optimized parameters of the two SVLS gratings are summarized in Table~\ref{tab:svls_grating_parameters}. As of December 2025, only the medium-resolution SVLS grating has been installed and commissioned.

\begin{figure}[t]
  \centering
  \includegraphics[width=0.95\linewidth]{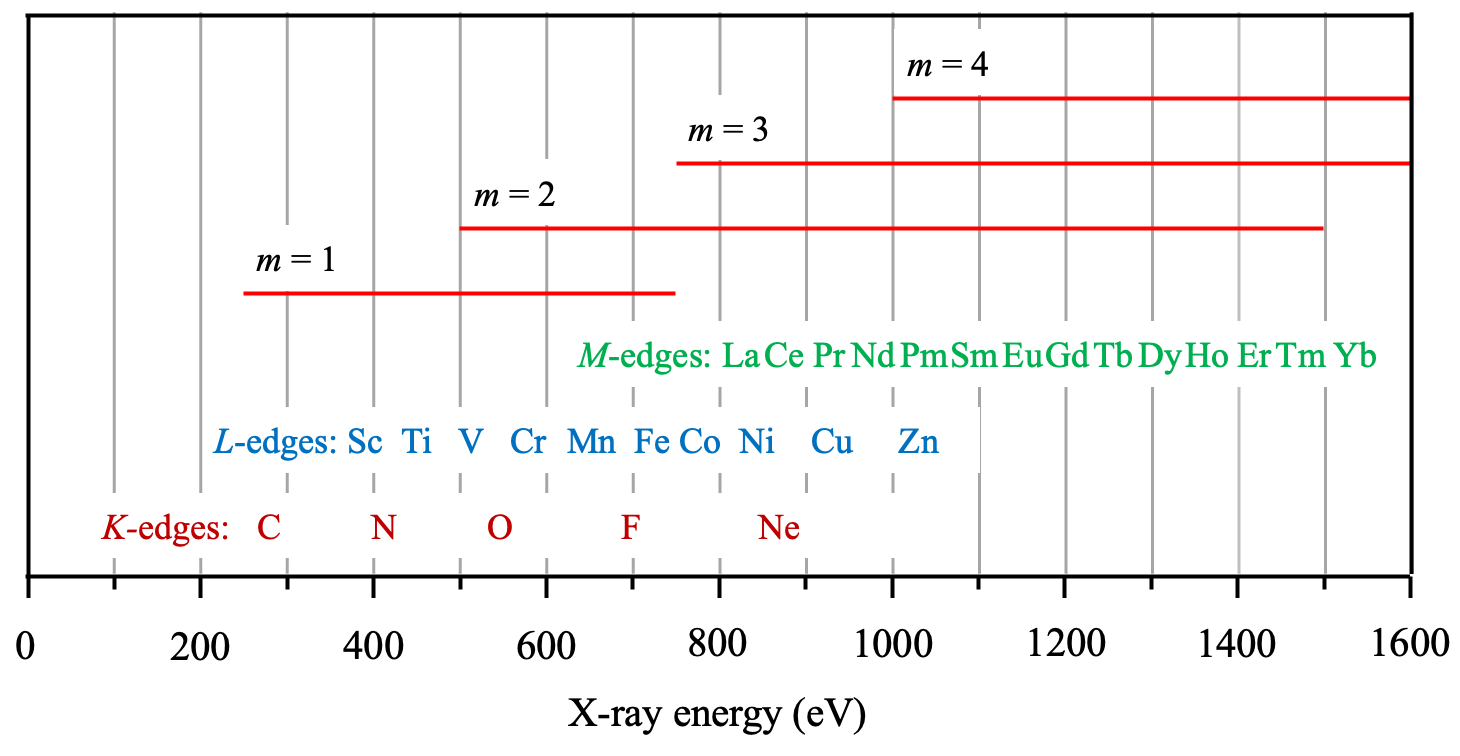}
  \caption{Accessible photon-energy ranges and diffraction orders of the qRIXS spectrometer, together with representative elemental absorption edges within the design window.}
  \label{fig:svls_energy_range}
\end{figure}

The accessible photon-energy ranges and diffraction orders of the optimized spectrometer are shown in Fig.~\ref{fig:svls_energy_range}. The first-order range spans 250--750~eV, and all 3$d$ transition-metal $L$ edges from V upward can be reached in at least two diffraction orders, allowing throughput to be traded against resolving power.

\subsubsection{qRIXS detectors}
LCLS utilizes advanced high-speed data acquisition systems to meet the demands of ultrafast and high-repetition-rate X-ray experiments. At qRIXS, Wave8 digitizers are used for pulse-resolved readout of intensity diagnostics, such as photodiodes and fluorescence intensity monitors (FIMs), enabling reliable shot-to-shot measurements of X-ray pulse energy~\cite{Wave8_LCLS}. In parallel, High-Speed Digitizers (HSDs) provide broadband waveform acquisition with 6.4 GHz sampling rates, allowing detailed characterization of detector signals and timing tools~\cite{HSD_LCLS}. These systems are integral to the LCLS diagnostics framework and are part of the broader data acquisition infrastructure developed for LCLS-II~\cite{Carbajo2018LCLSDAQ}.
The qRIXS endstation detector readout is based on the Archon CCD controller developed by Semiconductor Technology Associates (STA)~\cite{archon_sta}. The Archon system provides a modular and highly configurable readout platform designed for large-format scientific CCD sensors requiring low-noise performance and flexible clocking architectures.
For operation at LCLS, the detector system is fully integrated into the LCLS-II Data Acquisition (DAQ) pipeline~\cite{lcls2daq}. This integration enables deterministic synchronization between detector acquisition, accelerator timing signals, and experiment control systems. Detector exposure and readout cycles are synchronized with the facility timing system which ensures reproducible correlation between recorded detector frames and individual X-ray pulse events.
\paragraph{Detector Configuration}

The qRIXS detector employs a CCD sensor with \(4224 \times 1024\) pixels, with 4224 pixels of 10~\(\mu\mathrm{m}\) pitch along the spectrometer's dispersive direction and 1024 pixels of 20~\(\mu\mathrm{m}\) pitch along the non-dispersive direction. The finer pitch along the dispersive direction provides better sampling of the energy spectrum, while the full sensor is \(42 \times 20\)~mm\(^2\). The Archon controller supports flexible readout configurations for both one-dimensional (1D) and two-dimensional (2D) acquisition.

\begin{figure*}[t]
  \centering
  \includegraphics[width=0.99\linewidth]{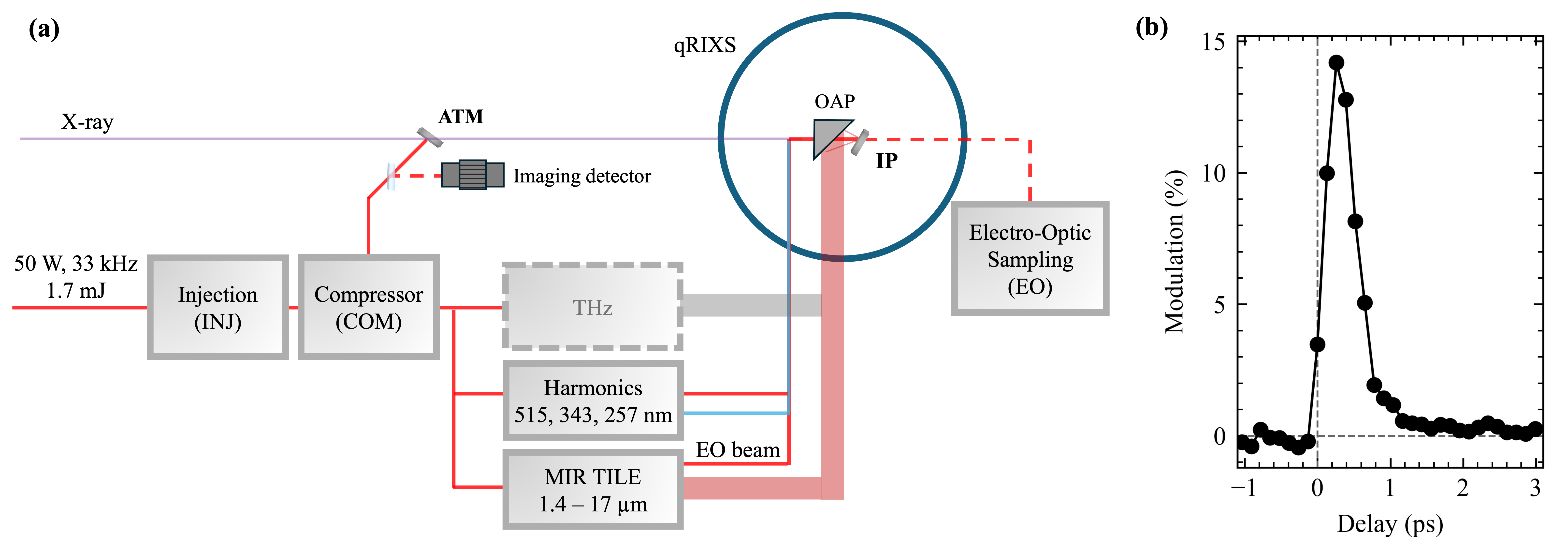}
\caption{(a) Schematic of the optical-laser system and beam delivery to qRIXS. Stretched 1030~nm pulses from the NEH Laser Hall are aligned and resized in the injection (INJ) TILE and compressed to near the transform limit in the compressor (COM) TILE. A fraction is sent to the arrival-time monitor (ATM), where the optical and X-ray beams cross at \(20^\circ\) for spatially encoded time-of-arrival measurements. The main beam is delivered to the interaction point (IP) directly or after harmonic generation, optical parametric amplification and difference-frequency generation, or THz generation. To focus the fundamental and harmonics beams at the IP we use a lens with a long Effective Focal Length (EFL) (> 1 m) located outside of vacuum, while for MIR and THz an in-vacuum hexapod-controlled Off-Axis Parabola (OAP) with EFL = 76 mm is used. Electro-optic (EO) sampling characterizes the mid-infrared (MIR) and THz pulses at the IP. (b) EO sampling trace in ZnTe for a 15~$\mu$m MIR pump with a 1030~nm probe and an unbalanced single-detector readout, showing the MIR pulse envelope and a slowly decaying pump-induced response in ZnTe.}
  \label{fig:las_1}
\end{figure*}

\paragraph{One-Dimensional High-Rate Readout}
During standard qRIXS operation, the detector is read out in a one-dimensional configuration in which the 1024 pixels along the non-dispersive direction of the CCD are summed on-chip into a single row, yielding an effective image of $4224 \times 1$ pixels along the dispersive direction. Because the horizontally collimating collection mirror suppresses the curvature of the spectral lines across the detector (Sec.~\ref{sec:optic_scheme}), this binning does not degrade the energy resolution. Binning improves signal-to-noise performance through charge integration along the non-dispersive direction while significantly reducing readout time and data throughput requirements. This configuration enables acquisition rates of up to 102\,Hz.
\paragraph{Two-Dimensional Acquisition Mode}
For detector diagnostics, alignment procedures, and spatial characterization measurements, the system can be operated in  2D-mode utilizing the full $4224 \times 1024$ pixel array. Preservation of spatial information in both detector dimensions allows verification of spectrometer focus and detector response uniformity.
Due to increased readout bandwidth and data volume, the achievable acquisition rate in 2D-mode is approximately 1\,Hz.
\paragraph{LCLS-II DAQ Integration}
Detector data generated by the Archon controller are streamed directly into the LCLS-II DAQ system, where event building, timestamp association, and metadata correlation are performed in real time. Synchronization with the accelerator timing system guarantees deterministic association between detector frames and experimental parameters such as beam energy, machine state, and scan variables.
\paragraph{Custom EPICS IOC for Detector Control}
Detector operation and experiment-level control are implemented through a custom-built EPICS Input/Output Controller (IOC). The IOC provides an abstraction layer between the Archon hardware interface and the LCLS control environment which enables standardized interaction with the detector through EPICS process variables (PVs).
The custom IOC manages detector configuration parameters including acquisition mode selection, exposure timing, binning configuration, triggering control, and readout state monitoring. Communication between EPICS and the Archon controller is implemented via network-based command and status interfaces.
Through EPICS integration, detector operation can be synchronized with motion control systems, scan sequences, and facility timing triggers. This control approach enables automated data acquisition procedures and reproducible experimental operation within the qRIXS endstation.

\section{\label{Laser}Optical Lasers}

\begin{table*}
\caption{\label{tab:las1}qRIXS laser parameters estimated at the interaction point for the pump wavelengths used during the first year of operation. All values are estimated at the IP for a 33~kHz repetition rate. Spot sizes are FWHM obtained from gaussian fitting of the laser profiles from imaging cameras, and peak fluence is calculated as $2E/(\pi w^2)$, where $w$ is the $1/e^2$ radius and $E$ is the energy per pulse.}
\begin{ruledtabular}
\begin{tabular}{lccccc}
Parameter & 1030 nm\footnote{Generated with spectrally broadened pulses via the multi-pass cell (MPC).} & 515 nm\footnotemark[1] & 257 nm\footnotemark[1] & 5 $\mu$m\footnote{Generated with native Carbide pulses.} & 15 $\mu$m\footnotemark[2] \\
\hline
Pulse duration (fs)             & 30    & 30   & 30  & 300 & 300 \\
Max pulse energy ($\mu$J)       & 300   & 50   & 5   & 5   & 0.3 \\
Min spot size ($\mu$m)          & 250   & 180  & 140 & 80  & 100 \\
Max peak fluence (mJ/cm$^2$)         & 424   & 136  & 22  & 69 & 2.6   \\
Max average power (mW)          & 10000\footnote{Limited to 10~W to avoid thermal damage of the in-vacuum in-coupling mirror.} & 1650 & 165 & 165 & 10 \\
\end{tabular}
\end{ruledtabular}
\end{table*}

\subsection{Laser system and delivery}
The qRIXS instrument utilizes optical laser pulses generated in the nearby NEH Laser Hall Laboratory with a Carbide CB3-80W laser system from Light Conversion. The laser is seeded by a Light Conversion Flint oscillator laser which is time-locked to the RF frequency from the LCLS-II timing system. The Carbide source generates pulses of 2~mJ energy, centered at 1030 nm wavelength with a native bandwidth of 7.5~nm and a pulse duration of 330 fs. The system operates at a variable repetition rate, typically set to 33 kHz during the first phase of commissioning of the qRIXS instrument. 

The native Carbide laser beam can be spectrally broadened in a Multi-Pass Cell (MPC) resulting in a bandwidth of up to 100~nm, capable of sustaining pulses as short as 25~fs. Such technique has been demonstrated extensively in the laboratory environment \cite{schulte2016nonlinear-a3c,hariton2023spectral-884, tavella2022multipass-d78} and was here implemented for the first time at LCLS, achieving excellent beam quality and stability for operational needs. 

The laser beam – either the native Carbide or the MPC spectrally broadened pulses – is delivered to an optical table in the qRIXS hutch to achieve proper beam conditioning for qRIXS experiments. Figure \ref{fig:las_1}a shows the schematic layout of the beam delivery to the instrument. Beam condition is achieved in modular setups (called TILEs), including  wavelength conversion units generating harmonics (HRM), tunable IR and MIR beams (OPA, DFG), and THz beams. The main laser parameters are listed in Table \ref{tab:las1}. 

\begin{figure*}[t]
  \centering
  \includegraphics[width=0.99\linewidth]{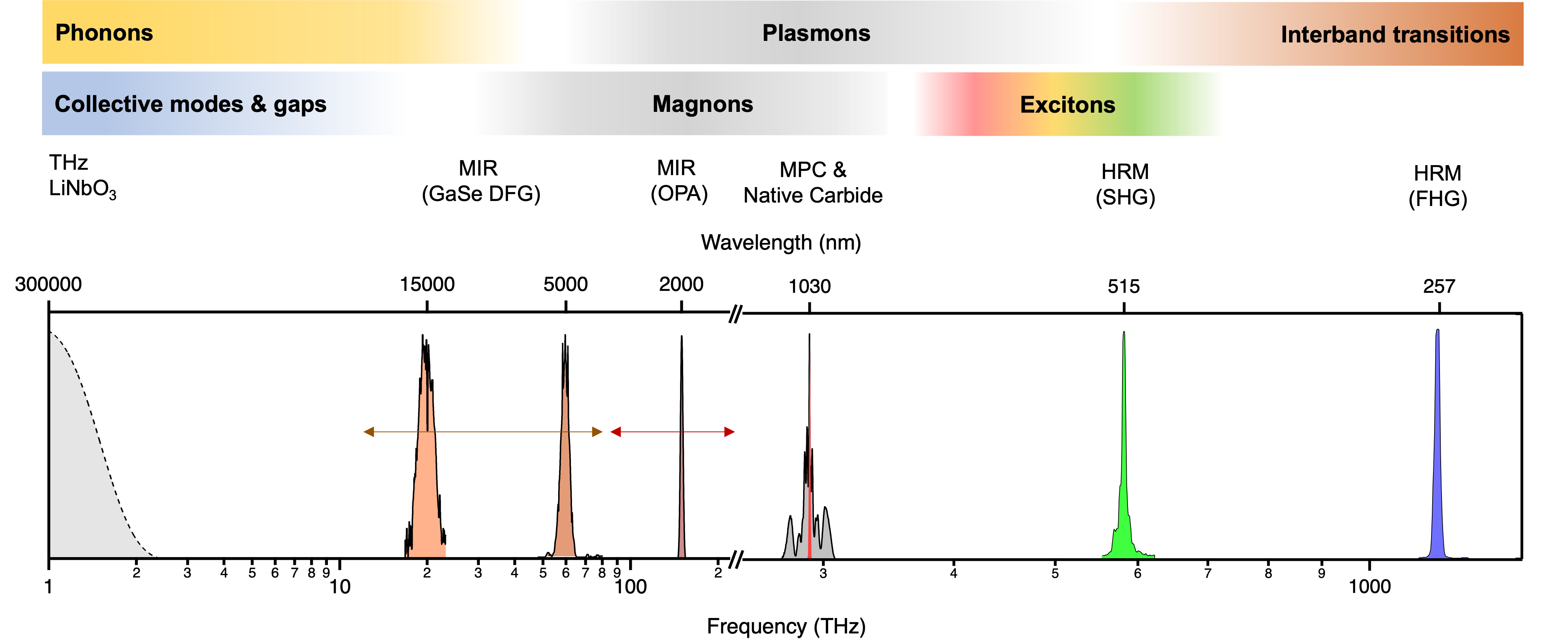}
  \caption{Spectra of the laser pulses available at qRIXS IP and corresponding energy ranges of the relevant excitations in materials. Spectra are measured experimentally for DFG (light brown shading), MPC (solid line, gray shading), SHG (green shading) and FHG (purple shading) cases, and estimated for OPA (brown shading), native Carbide (magenta shading) and THz (dashed line, gray shading). Arrows indicate the spectral regions where the pulses are tunable in wavelength.}
  \label{fig:las_2}
\end{figure*}

In Fig.~\ref{fig:las_2} we also display how the driving laser spectra match with the relevant excitations for material science experiments at qRIXS. Excitons and interband transitions, such as charge transfers and d-d excitations, can be photo-excited with harmonics of the fundamental, namely second harmonic generation (SHG) at 515 nm and fourth harmonic generation (FHG) at 257 nm. Below charge transfer edges, more typical non-selective electronic and above-bandgap excitations can achieved with 1$\mu$m excitation, generally reproducing results from commonly used 800 nm excitations \cite{Bostedt_rmp_2016,mitrano-sa-2019,wandel-science-2022}. 

Optical Parametric Generation (OPA) allows the pump to span a broad spectral range for selective and narrow-band excitation of magnons, plasmons and IR-active phonons. These tunable NIR and MIR beams are generated with an Orpheus-ONE-HE from Light Conversion. The output covers a range from 1.3 to 17 um with varying levels of power and laser parameters. The two MIR wavelengths used during the first year of operation at qRIXS are shown in Table \ref{tab:las1}, respectively 5 and 15 $\mu$m at 33~kHz. Further development work is ongoing to boost the available fluence at the long wavelength end of this regime and extend operations down to 20 $\mu$m. 

THz capabilities are currently under commissioning and allow covering lower energy excitations, such as soft phonon modes and other low-lying collective modes. The THz generation scheme follows previous development at the LCLS using pulse-front tilt THz generation in LiNbO$_3$ \cite{hoffmann2007efficient-fea,kramer2020enabling-abc}. With this scheme we expect to generate fields on the order of 150 kV/cm at 33 kHz. 

These laser beams are in-coupled through distinct paths into the qRIXS chamber. MIR and THz beams are focused to the IP by a parabolic mirror with a 3-inch (76.2~mm) effective focal length, located inside the qRIXS chamber (see Section~\ref{qRIX} and Fig.~\ref{fig:las_1}). The fundamental beam, as well as its harmonics, are delivered via rectangular UV-enhanced aluminum mirror upstream of the path of the x-ray beam, reflecting the beam onto the IP close to a collinear geometry (<1 deg). To avoid damage to the in-vacuum mirror the average power is typically limited to 10 W, thus limiting the maximum fluence at IP for the fundamental wavelength. In case of need of higher fluence a lower repetition rate can be used. 

A small fraction of the beam can be used for Electro-Optic (EO) sampling purposes to characterize the MIR and the THz beams in terms of field and envelopes at the IP location Fig. ~\ref{fig:las_1}b. The EO sampling line is also used to characterize the time of arrival of both MIR and THz laser beams onto the IP with respect to the 1~$\mu$m fundamental, which can be more easily co-timed to the x-ray beam (see section \ref{OP}). 

\subsection{Arrival Time Monitor}

The pump laser system is synchronized to the FEL x-ray pulse via RF signal locking. This FEL synchronization method carries inherent time-of-arrival uncertainty due to many factors, as extensively reported in the literature \cite{krupin2012temporal-5dd,beye2012xray-9d6, bionta2014spectral-680, eckert2015principles-2ec, Droste20}. To measure the shot-to-shot x-ray time-of-arrival instability with respect to the pump laser, an Arrival Time Monitor (ATM) is necessary. For this purpose, a fraction of the fundamental laser beam available is split off and delivered to the ATM, a separate interaction chamber for laser and x-ray beams. 

In this chamber we perform X-ray pump, laser reflectivity probe measurements on specially designed GaAs-based thin-film targets to characterize the time of interaction between the two pulses with femtosecond precision on a shot-to-shot basis, similarly to what has previously been done at LCLS and other FEL facilities\cite{krupin2012temporal-5dd, beye2012xray-9d6, bionta2014spectral-680, eckert2015principles-2ec, Droste20}. The qRIXS ATM operates in a spatial-encoding geometry: the target surface is tilted by \(20^\circ\) with respect to the X-ray beam while the laser probe strikes it at normal incidence, so that the two beams cross at \(20^\circ\) on the target surface (Fig.~\ref{fig:las_1}). The X-ray arrival time therefore varies linearly along the X-ray footprint, and imaging the footprint maps the arrival time onto position on the camera.

The ATM target is oriented to reflect the laser pulse at normal incidence, and the reflected beam is imaged onto a camera (Teledyne Adimec Opal) for differential analysis between ON (on-time) and OFF shots (delayed by tens of ns, so effectively free of pump-induced effects). When the X-rays are absorbed by the target, its reflectivity changes due to the carriers excited into the conduction band. To maximize this optical response, the target is an optimized stack of thin semiconducting films with the active GaAs layer on top. The stack acts as an etalon that boosts the reflectivity change by orders of magnitude, as discussed in the literature\cite{eckert2015principles-2ec, Droste20, Muhammad-cleo-2021}. Details on the target design and performance will be reported in a future publication.

\begin{figure}[tb]
  \centering
  \includegraphics[width=0.98\linewidth]{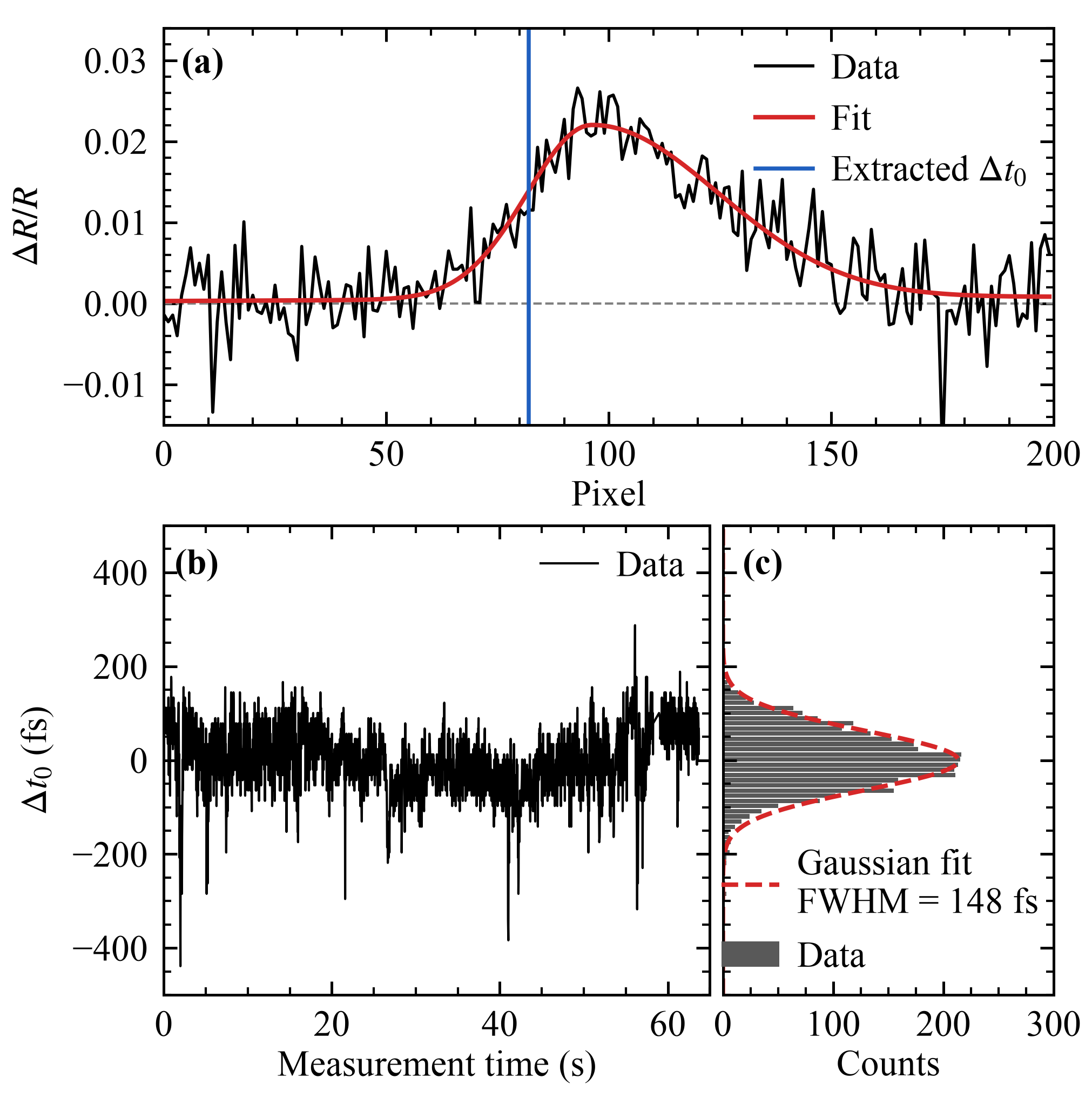}
  \caption{ATM performance during initial commissioning of qRIXS. (a) Typical pump-probe profile from a single detector exposure (black) with a fit to exponential rise and decay functions (red). The vertical blue line marks the extracted time of arrival $\Delta t_0$. (b) Time of arrival as a function of measurement time. (c) Histogram of the arrival times in (b), with one bin per detector pixel (11~fs). The dashed red line is a Gaussian fit with a FWHM of 148~fs.}
  \label{fig:las_3}
\end{figure}

The ATM diagnostic measurement is typically performed in parallel to the ongoing experiment at the IP in qRIXS to provide real-time correction for timing jitter and drift. In this configuration the ATM target intercepts only the edge of the X-ray beam, typically a few tens of percent of the pulse energy, while the remainder propagates to the IP. However, in the case of low X-ray flux such as for the case of high-resolution RIXS the ATM can be operated sequentially for periodic checks of the timing drift.

Because of the combination of high repetition rate, low x-ray fluence and the slow integrating detectors utilized in qRIXS, the qRIXS ATM is not designed to perform shot-to-shot characterization at the full 33 kHz, rather it is designed to provide the average time-of-arrival for each single detector exposure of the RIXS spectrometer, i.e, at 102 Hz.

A typical trace recorded by the ATM is shown in Fig.~\ref{fig:las_3}a. The edge indicates the sudden jump in reflectivity reflecting the time of arrival of the x-ray in respect to the laser beam. The resolution of the ATM is given by the 20° angle of incidence and the pixel size, giving a $\approx$2 fs accuracy in the edge location. Further timing uncertainty derives from the signal-to-noise in the ATM trace and from relative drift and jitter between the two laser paths (to the IP and to the ATM). The total time window of the ATM is limited by the width of the X-ray spot on the target (which is constrained by focusing requirements at the experimental IP), but is typically about 1 ps. To maintain high quality time resolution across the entire time range (and to account for changes in the laser pathlength to the IP between experiments), the ATM is also equipped with a mechanical delay stage which compensates for changes in the laser X-ray timing at the IP such that the timing signal stays centered on the ATM, even for delay scans beyond the ATM temporal window.

Fig.~\ref{fig:las_3} also shows the typical jitter and drift levels during few minutes of operations. The shot-to-shot jitter of the time-of-arrival data averaged over the 102 Hz bins of integration time of the detector is shown in Fig. \ref{fig:las_3}b and has a FWHM of $\approx$150 fs. Slower drift can be as high as 0.2 ps/min, and thus pose a serious challenge for experiments in which the ATM cannot be operated simultaneously with the experiment. This type of situation requires periodic checks of the time of arrival. Further development work is ongoing to i) improve the performance of the ATM, ii) improve the available flux for monochromatized X-ray at both ATM and qRIXS, and iii) improve the timing stability at the LCLS.

\section{\label{OP}Instrument Operation and Performance}
In this section, we will present the performances of our incident flux monitor (FIMs), fine timing tool at the IP, and RIXS spectrometer. We will also briefly describe the protocol of typical time-resolved RIXS measurements at qRIXS.
\subsection{Timing}

\begin{figure}[b]
  \centering
  \includegraphics[width=0.99\linewidth]{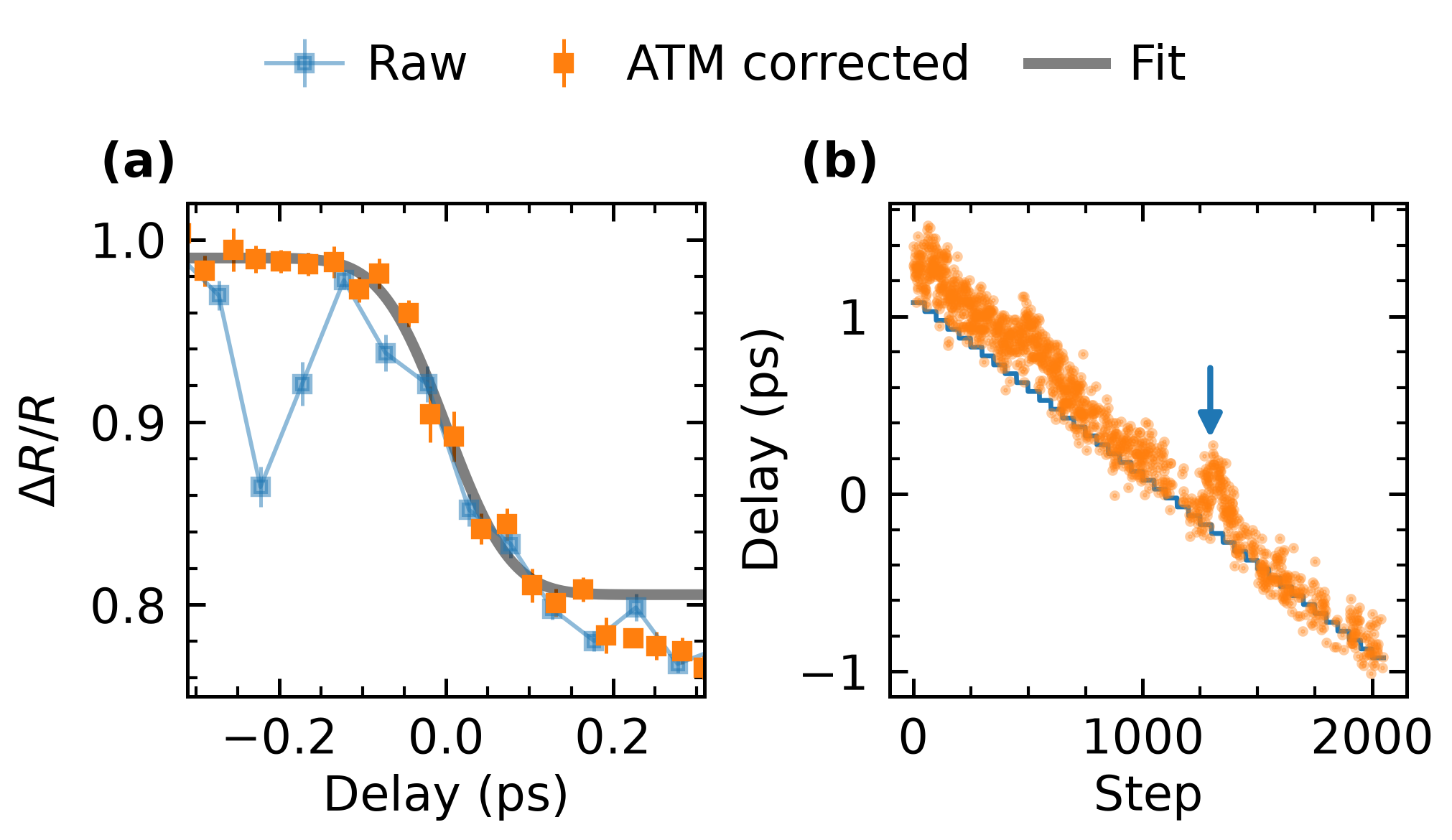}
  \caption{(a) Fine-timing scans measured at the interaction point before and after ATM correction. (b) Pump laser delay value tracked by the ATM over the course of the measurement. The blue arrow marks a discrete timing jump, which gives rise to the intensity discontinuity near -0.1 ps in panel (a).}
  \label{fig:6}
\end{figure}

As discussed in Section~\ref{Laser}, a dedicated timing sample is available at the interaction point for synchronization of the pump and X-ray pulses. GaAs is used for the 1030 nm fundamental and its second harmonic, while SiNx is used for the fourth harmonic \cite{Krupin-oe-2012,Muhammad-cleo-2021}. For mid-infrared pumps, synchronization is achieved through a cross-correlation scheme in which the X-rays are first synchronized to the native Carbide laser (1030~nm), followed by synchronization of the Carbide laser to the mid-infrared pump using ZnTe. 

Figure~\ref{fig:6} illustrates this procedure for 1030 nm (MPC) operation. The measurements were carried out with the XFEL lasing at 102 Hz and with the undispersed SASE beam (zero order configuration of the monochromator). During the measurement, the qRIXS ATM continuously tracked the X-ray arrival time relative to the pump laser. A clear anomaly is visible near − 0.1~ps in the raw data [Fig.~\ref{fig:6}(a)], which is traced to a timing jump revealed by the ATM-corrected delay [Fig.~\ref{fig:6}(b)]. Applying the ATM correction restores a smooth decay and yields a time resolution of 0.146(15) ps.

\subsection{RIXS spectrometer}
\subsubsection{Resolving power}
\begin{figure}[h]
  \centering
  \includegraphics[width=0.85\linewidth]{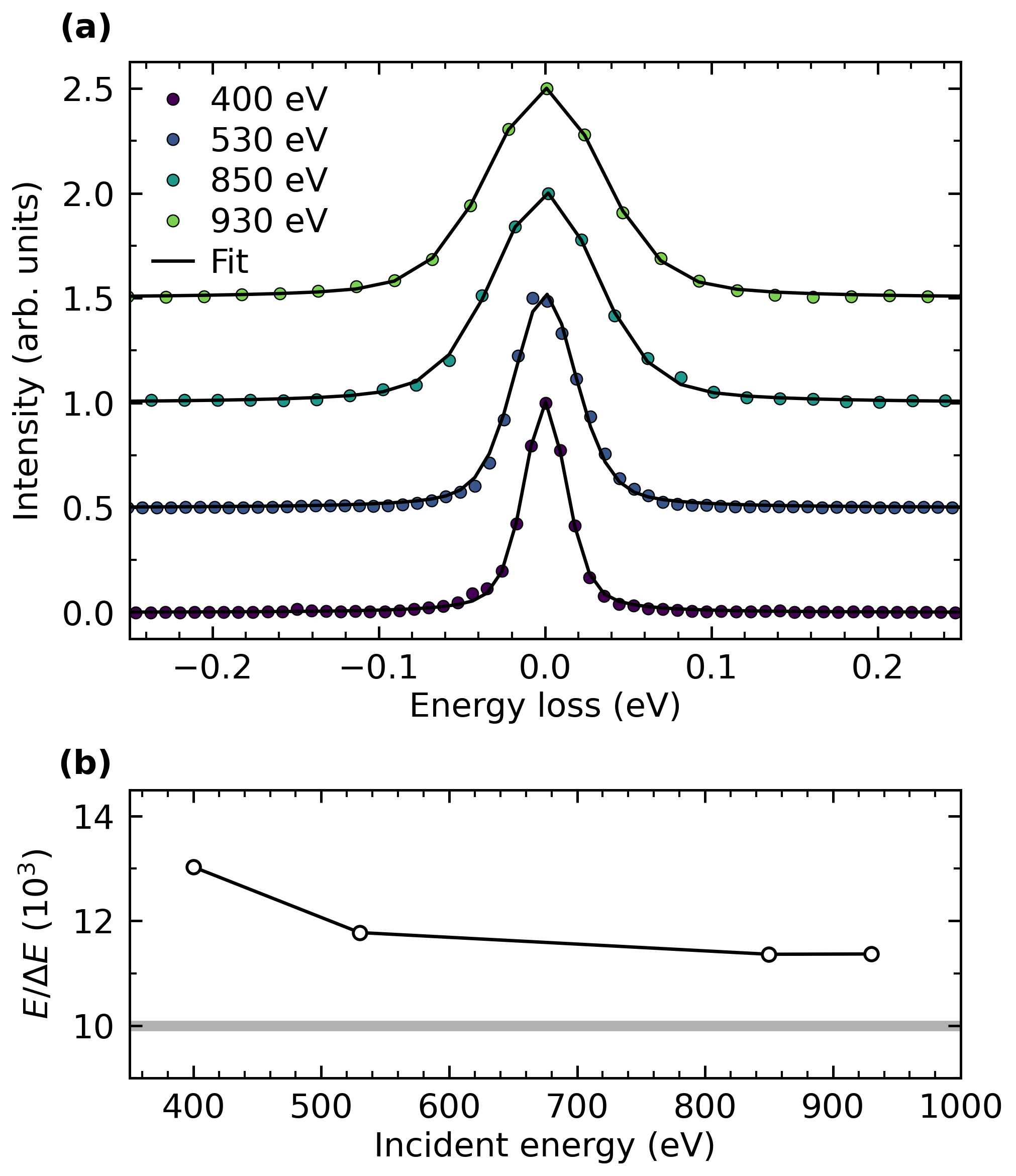}
  \caption{(a) Elastic line measurements of carbon tape (b). Combined resolving power as a function of incident X-ray energy.}
  \label{fig:7}
\end{figure}

The combined resolution of the RIXS spectrometer, operating under the 1500~l/mm SVLS grating, is measured by the carbon tape. The elastic lines at various soft X-ray resonant energies are shown in Fig.~\ref{fig:7}. The profile remains sharp and symmetric at all X-ray energies measured and can be well reproduced by a Gaussian function. The corresponding resolving power (RP) at all energies measured remains above 10$^{4}$, matching the designed performance of the relevant optics. 

\subsubsection{Oxygen \textit{K}-edge}

\begin{figure}[t]
  \centering
  \includegraphics[width=0.95\linewidth]{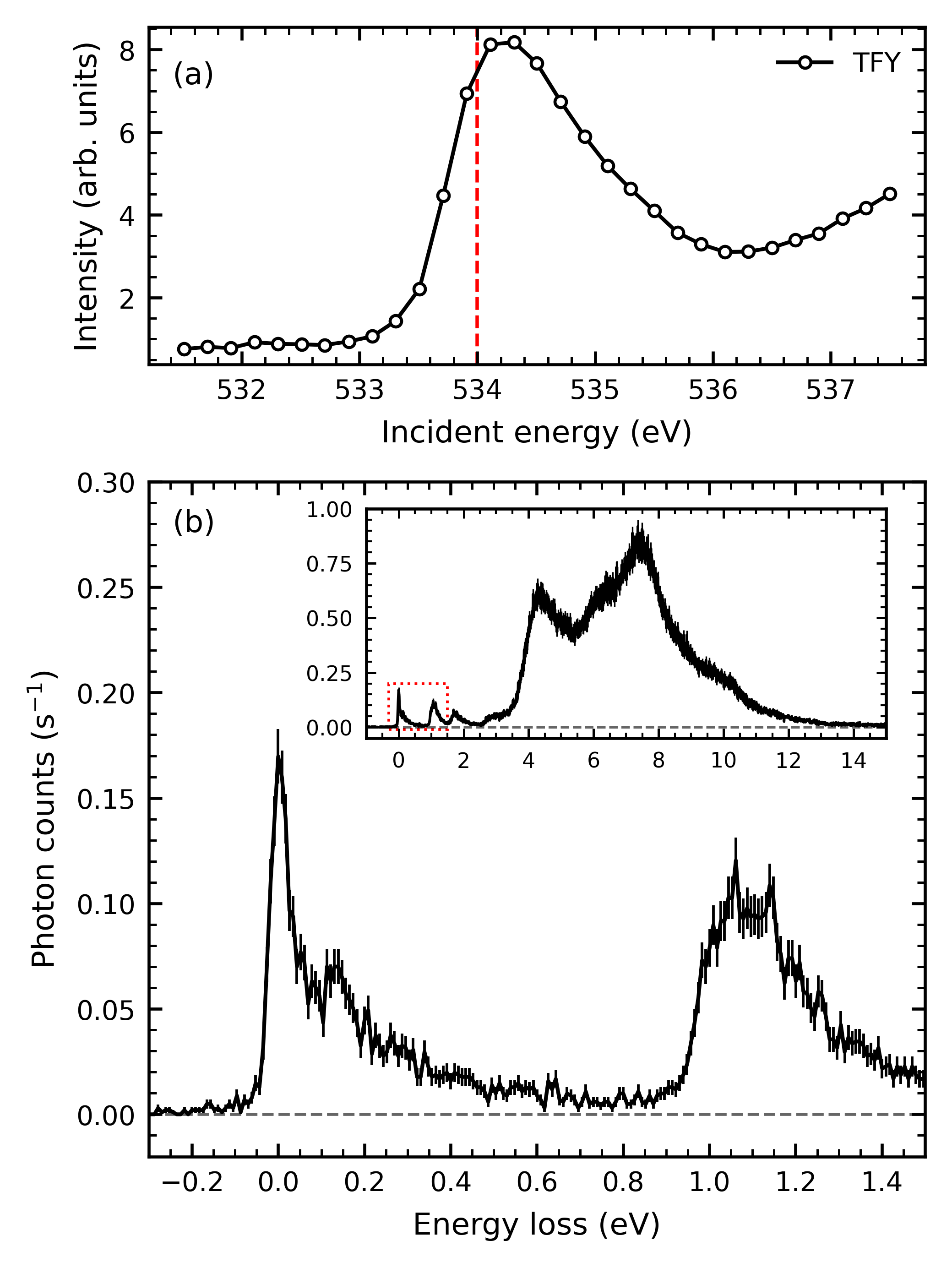}
  \caption{(a) Total fluorescence yield (TFY) X-ray absorption spectrum of NiO near the O \(\mathit{K}\) edge. (b) Expanded view of the RIXS excitations near the elastic line; the inset shows the full RIXS spectrum. The spectrum was acquired at an incident X-ray energy of 534~eV, as indicated by the red line in panel (a). Error bars are one standard deviation Poisson counting uncertainties.}
  \label{fig:8}
\end{figure}

We use NiO to demonstrate the performance of the RIXS spectrometer at the O \textit{K} edge. Figure~\ref{fig:8} shows the O $\mathit{K}$-edge RIXS response of NiO measured with qRIXS. Figure~\ref{fig:8}(a) presents the total-fluorescence-yield X-ray absorption spectrum across the O $\mathit{K}$ edge, with the incident photon energy for the RIXS measurement set to 534~eV, as indicated by the red line. The corresponding RIXS spectrum, measured at diffractometer angle $\theta = 45^{\circ}$ and spectrometer arm angle $2\theta = 145.58^{\circ}$, is shown in Figs.~\ref{fig:8}(b). The overall spectral lineshape reproduces the major excitation features previously reported in O \textit{K}-edge RIXS studies of NiO~\cite{Duda-prl-2006}. In particular, the full spectrum exhibits the broad charge-transfer manifold above 5~eV energy loss together with a shoulder around 4--5~eV, consistent with the previously reported nonlocal charge-transfer excitation. At lower energy loss, the spectrum resolves an intense feature near 1.1~eV and additional spectral weight around 1.7~eV, consistent with the \(dd\) excitations and double-singlet creation (DSC) discussed in the earlier work. The overall energy resolution of approximately 0.05~eV represents an order-of-magnitude improvement over the 0.5~eV resolution of that study, allowing clearer observation of the low-energy spectral structure, including spectral weight close to the elastic line.

\subsubsection{Copper \textit{L}-edge}
\begin{figure}[t]
  \centering
  \includegraphics[width=0.95\linewidth]{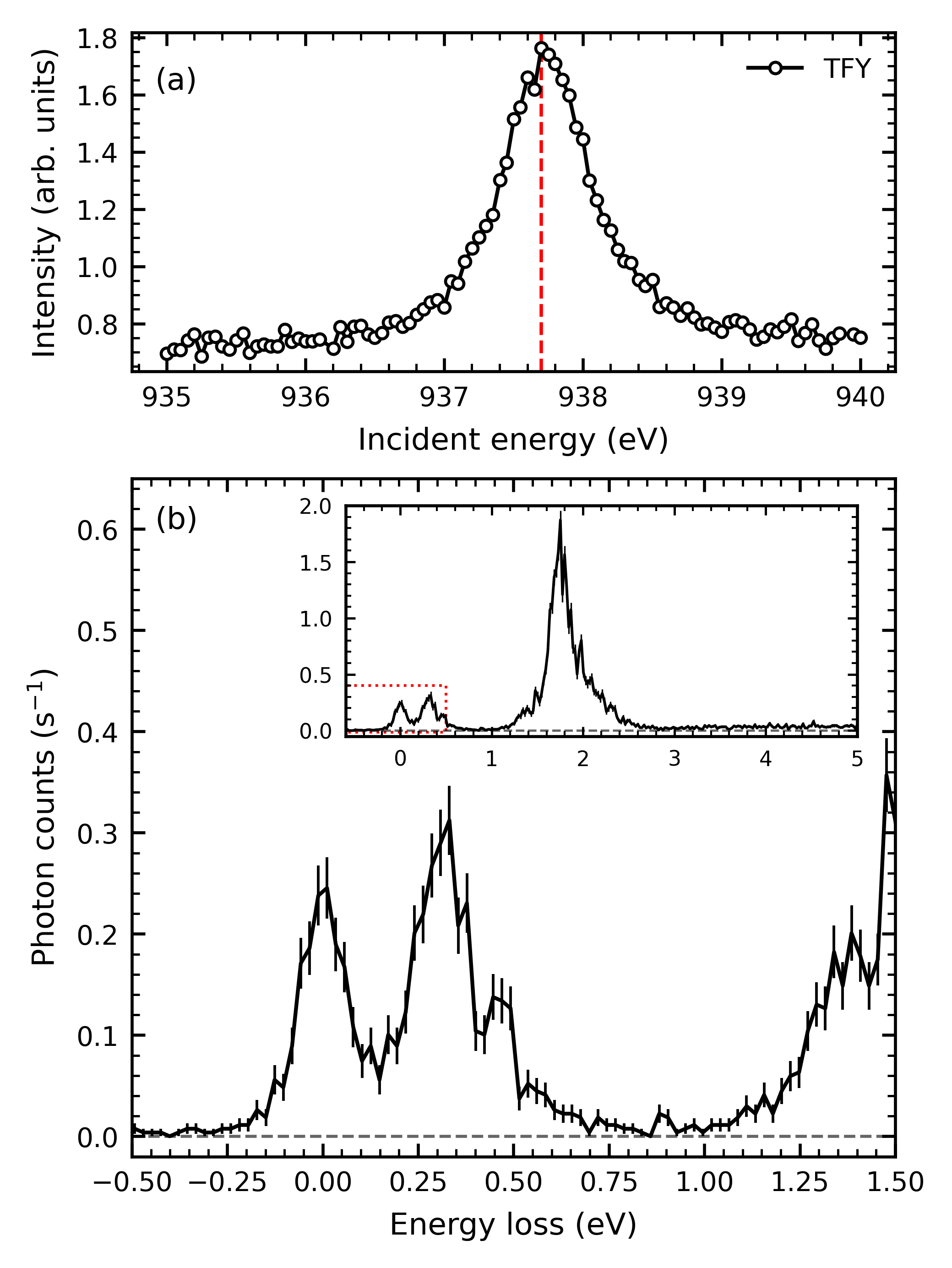}
 \caption{Cu $L_3$-edge RIXS of La$_2$CuO$_4$. (a) Total fluorescence yield (TFY) X-ray absorption spectrum near the Cu $L_3$ edge. The red line marks the incident photon energy of 937.7~eV~\cite{CuLnote} used for the RIXS measurement. (b) RIXS spectrum near the elastic line, acquired at the momentum transfer $\mathbf{q} = (0.29, 0.29, 1.28)$ in reciprocal lattice units. The inset shows the full spectrum. Error bars are one standard deviation Poisson counting uncertainties. Measurement time was 270 seconds.}

  \label{fig:9}
\end{figure}

La$_2$CuO$_4$ is used here as a benchmark cuprate system to evaluate the qRIXS spectrometer performance at the Cu \textit{L} edge. The total-fluorescence-yield spectrum is shown in Fig.~\ref{fig:9}(a), and the incident photon energy selected for the RIXS measurement, 937.7~eV \cite{CuLnote}, is indicated by the red line. The resulting spectra, acquired at diffractometer angle $\theta = 25^{\circ}$ and spectrometer arm angle $2\theta = 145.58^{\circ}$, are displayed in Fig.~\ref{fig:9}(b).

Several characteristic features of the established Cu \textit{L}-edge RIXS response of La$_2$CuO$_4$ are clearly reproduced \cite{Robarts-prb-2021,Dean-NM-2013}. In the full spectrum, the intensity is dominated by the \textit{dd} excitation manifold between 1 and 3~eV, in agreement with previous Cu \textit{L}$_3$-edge studies. At higher energy loss, the spectrum develops a broad continuum extending roughly from 3 to 10~eV, which is consistent with charge-transfer excitations commonly observed in Cu \textit{L}$_3$-edge RIXS of cuprates. Closer to the elastic line, the higher resolution ($\sim$0.1~eV) of the present measurement resolves a quasielastic contribution together with a magnetic excitation near 0.32~eV and additional spectral weight centered around 0.47~eV. Earlier high-resolution work assigned this higher-energy component to structured multimagnon excitations. Overall, these data confirm that qRIXS reproduces the canonical Cu $L_3$-edge spectrum of La$_2$CuO$_4$, resolving the $dd$ excitations and the high-energy charge-transfer response and, most importantly, the low-energy magnetic excitations of this prototypical cuprate antiferromagnet, which were out of reach for the previous generation of soft X-ray spectrometers at LCLS~\cite{mitrano-sa-2019,Mitrano-prb-2019,paris-npjqm-2021,Parchenko-prr-2020,lu-sr-2020}.

\subsection{Time-resolved RIXS measurements}

\begin{figure}[t]
  \centering
  \includegraphics[width=0.95\linewidth]{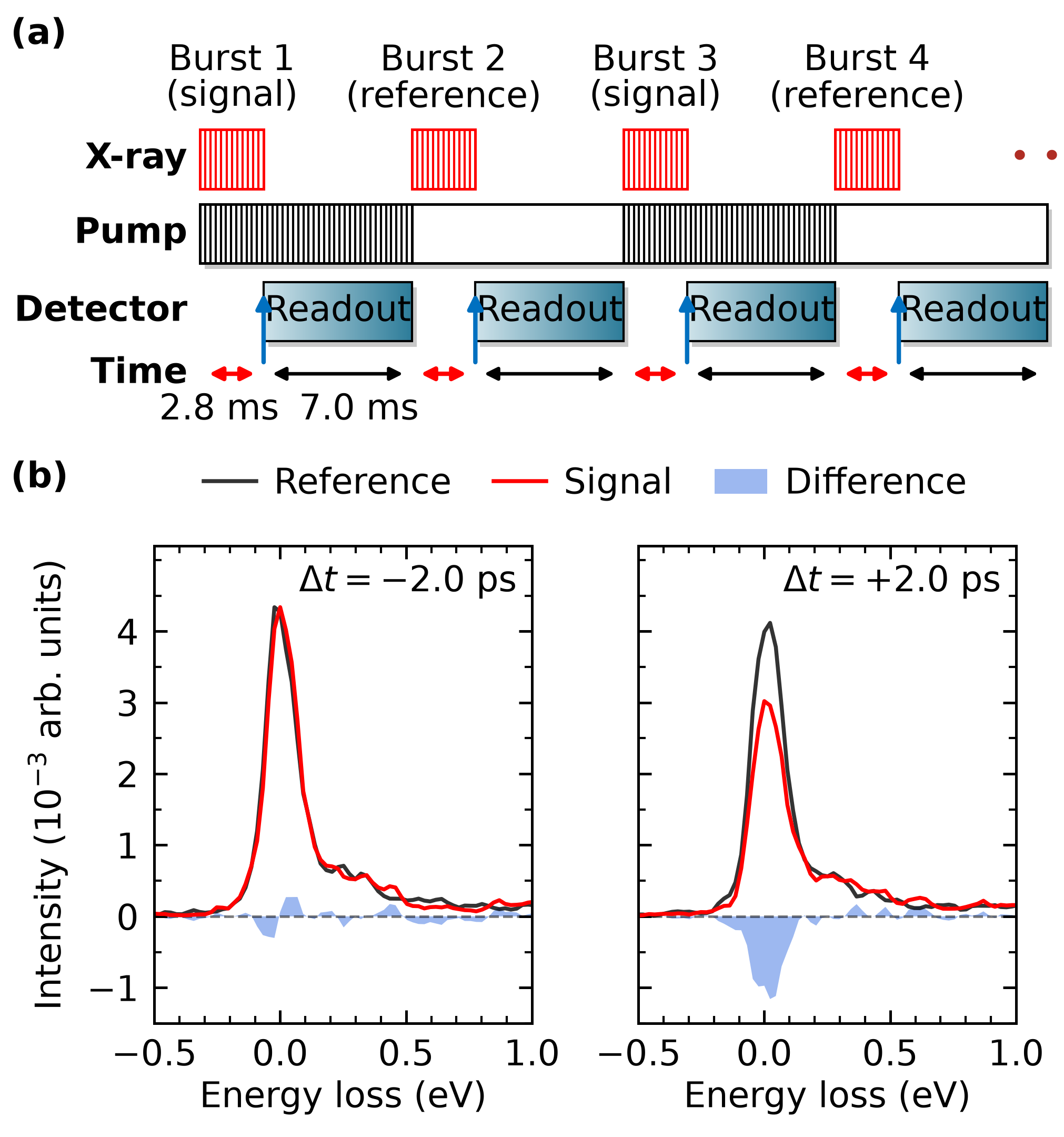}
  \caption{Time-resolved RIXS measurements. (a) X-ray burst scheme used to synchronize the pump laser, X-ray pulses, and integrating detector. Signal and reference frames alternate; the blue arrows mark the trigger that starts the detector readout. (b) Cu $L_3$-edge RIXS spectra of LBCO at $\Delta t = -2$~ps (left) and $+2$~ps (right) after excitation, measured at $\mathbf{q} = (0.23, 0, 1.72)$ in tetragonal reciprocal lattice units and $T = $25~K. Each pumped spectrum (signal) is shown with the unpumped reference spectrum recorded in the adjacent burst; the shaded area is their difference. The quasielastic line is suppressed after time zero, consistent with the photoinduced melting of charge-order correlations reported in Ref.~\onlinecite{mitrano-sa-2019}.}
  \label{fig:10}
\end{figure}

In this section, we describe the performance of time-resolved RIXS measurements at qRIXS. As discussed in Section~\ref{qRIX}, the RIXS detector (Archon) operates at a much lower frame rate (1 or 102~Hz) than the X-ray pulse rate (33~kHz). It therefore runs in an integrating mode, in which each detector frame accumulates signal from many X-ray pulses. Archon also has a finite readout interval, during which continued X-ray exposure can degrade the data quality. For this reason, we encourage users to perform time-resolved RIXS measurements in the so-called "bursting" mode whenever the cleanest possible data are required. A schematic of this operation mode is shown in Fig.~\ref{fig:10}(a). When Archon is operated in 1D mode at 102~Hz, the period between two consecutive frames is 9.8~ms and the detector readout time is approximately 7.0~ms. As a result, each detector frame can be exposed to X-rays for only about 2.8~ms, corresponding to a single X-ray burst, before the trigger initiates readout. To further improve the data quality, an unpumped reference frame can be recorded every other Archon frame. This reference serves as an internal baseline for the neighboring pumped frame and helps suppress high-frequency artifacts. In practice, this is achieved by turning off the pump laser every other 9.8~ms cycle.

Figure~\ref{fig:10}(b) shows representative results for the cuprate superconductor La$_{2-x}$Ba$_x$CuO$_4$ (LBCO), measured at the Cu \textit{L} edge using the bursting mode described above, with a combined energy resolution of approximately 0.1~eV. The data acquisition time is 10 seconds per delay. The pump wavelength and fluence were 1030~nm and 0.3~mJ/cm$^{2}$, respectively. We focus on the spectral region near the elastic line, where a previous tr-RIXS study revealed strong pump-induced melting of charge order \cite{mitrano-sa-2019}. The reference spectrum is consistent with the spectrum measured before time zero ($-2.0$~ps), confirming that both represent the equilibrium state. In addition, a broad shoulder is resolved between 0.2~eV and 0.5~eV, consistent with magnon excitations \cite{Hu-pnas-2017}. Bursting operation inevitably reduces the duty cycle. When the time-resolved RIXS response is sufficiently strong and high-frequency artifacts are not a limiting factor, the reference frames are not required, and measurements can instead be performed in continuous mode.

\section{\label{conclusion}Outlook and Summary}

Since the inauguration of the LCLS-II linac in September 2023, qRIXS has undergone approximately six months of commissioning, followed by multiple user experiments over two LCLS runs. The delivered performance is in good agreement with the Phase~1 design goals, including a resolving power exceeding 10,000, operation at 33~kHz, and access to a broad pump-laser wavelength range spanning the ultraviolet to the mid-infrared. At the same time, several additional capabilities are planned. These include deployment of the blazed high-resolution monochromator gratings HEG and LEG, which are expected to further improve X-ray transmission efficiency (Table~\ref{tab:2}), an ultrahigh-resolution spectrometer grating with 3600~l/mm, and further improvement of the wavelength range and fluence in the long-wavelength MIR region, as well as full commissioning of the THz pumping capability. We plan to commission Photoelectron Spectrometry for Analysis of X-rays (PAX, Fig.~\ref{fig:1}) to monitor incident X-ray energy drifts on timescales of seconds to minutes and provide feedback for energy stabilization~\cite{Higley-jsr-2022}. Looking further ahead, installation of a Delta undulator would provide arbitrary control of the incident X-ray polarization, while a multi-layer polarimeter on the spectrometer arm would enable linear polarization analysis of the scattered photons.

In summary, we have described the principal components of qRIXS and the upstream beamline systems required for its operation. Commissioning measurements demonstrate high-resolution RIXS across a broad soft X-ray energy range, synchronized optical excitation from the ultraviolet to the mid-infrared, and burst-mode acquisition for time-resolved measurements with an integrating detector. These results establish qRIXS as a powerful platform for high-repetition-rate, ultrafast, and high-energy-resolution RIXS studies of materials and their nonequilibrium dynamics.

\begin{acknowledgments}
This research was carried out at the Linac Coherent Light Source (LCLS) at the SLAC National Accelerator Laboratory. SLAC National Accelerator Laboratory is supported by the U.S. Department of Energy, Office of Science, Office of Basic Energy Sciences under Contract No. DE-AC0276SF00515. The authors thank Giacomo Ghiringhelli, Daniele Di Castro, Giacomo Merzoni, Xi He, Ivan Bozovic, and Genda Gu for providing samples and insightful discussions. M.P.M.D's work at Brookhaven National Laboratory was supported by the U.S. Department of Energy (DOE), Division of Materials Science, under Contract No. DE-SC0012704. M.M. and S.F.R.T. are supported by the US Department of Energy, Office of Basic Energy Sciences, Early Career Award Program, under award no. DE-SC0022883. C.P. was supported by the Linac Coherent Light Source, SLAC National Accelerator Laboratory, under contract no. DE-AC02-76SF00515.
\end{acknowledgments}

\section*{Author Declarations}
\subsection*{Conflict of Interest}
The authors have no conflicts to disclose.

\section*{Data Availability}
The data that support the findings of this study are available from the corresponding author upon reasonable request.

\bibliography{qRIXS_REFs}% Produces the bibliography via BibTeX.

\end{document}